\documentclass[jair,twoside,11pt,theapa]{article}
\usepackage{jair, theapa, rawfonts}

\usepackage{graphicx}
\usepackage{wrapfig}
\usepackage[dvipsnames]{xcolor}
\usepackage{amsmath}
\usepackage{mathtools}
\usepackage{tikz}
\usetikzlibrary{decorations.pathreplacing}
\usetikzlibrary{fit,shapes}
\usepackage{pifont}
\usepackage{subcaption}
\usepackage{booktabs}
\usepackage{mdframed}
\usepackage{caption}
\RequirePackage[hyphens]{url}
\usepackage{url}
\usepackage{cleveref}

\newtheorem{proposition}{Proposition}
\newtheorem{corollary}{Corollary}
\newtheorem{definition}{Definition}

\jairheading{82}{2025}{471--501}{02/2024}{02/2025}
\ShortHeadings{AI Reliance \& Decision Quality}
{Schoeffer, Jakubik, Vössing, Kühl \& Satzger}
\firstpageno{471}

\begin{document}
\sloppy

\title{AI Reliance and Decision Quality: Fundamentals, Interdependence, and the Effects of Interventions}

\author{\name Jakob Schoeffer \email j.j.schoeffer@rug.nl \\
      \addr University of Groningen\\
      Groningen, The Netherlands
      \AND
      \name Johannes Jakubik \email johannes.jakubik@kit.edu \\
      \addr Karlsruhe Institute of Technology\\
      Karlsruhe, Germany
      \AND
      \name Michael Vössing \email michael.voessing@kit.edu \\
      \addr Karlsruhe Institute of Technology\\
      Karlsruhe, Germany
      \AND
      \name Niklas Kühl \email kuehl@uni-bayreuth.de \\
      \addr University of Bayreuth\\
      Fraunhofer FIT\\
      Bayreuth, Germany
      \AND
      \name Gerhard Satzger \email gerhard.satzger@kit.edu \\
     \addr Karlsruhe Institute of Technology\\
      Karlsruhe, Germany}

\maketitle

\begin{abstract}
\noindent
In AI-assisted decision-making, a central promise of having a human-in-the-loop is that they should be able to complement the AI system by overriding its wrong recommendations. In practice, however, we often see that humans cannot assess the correctness of AI recommendations and, as a result, adhere to wrong or override correct advice. Different ways of relying on AI recommendations have immediate, yet distinct, implications for decision quality. Unfortunately, reliance and decision quality are often inappropriately conflated in the current literature on AI-assisted decision-making. In this work, we disentangle and formalize the relationship between reliance and decision quality, and we characterize the conditions under which human-AI complementarity is achievable. To illustrate how reliance and decision quality relate to one another, we propose a visual framework and demonstrate its usefulness for interpreting empirical findings, including the effects of interventions like explanations. Overall, our research highlights the importance of distinguishing between reliance behavior and decision quality in AI-assisted decision-making.
\end{abstract}

\section{Introduction}\label{sec:introduction}

\begin{figure}[ht]
    \centering
    \includegraphics[width=\textwidth]{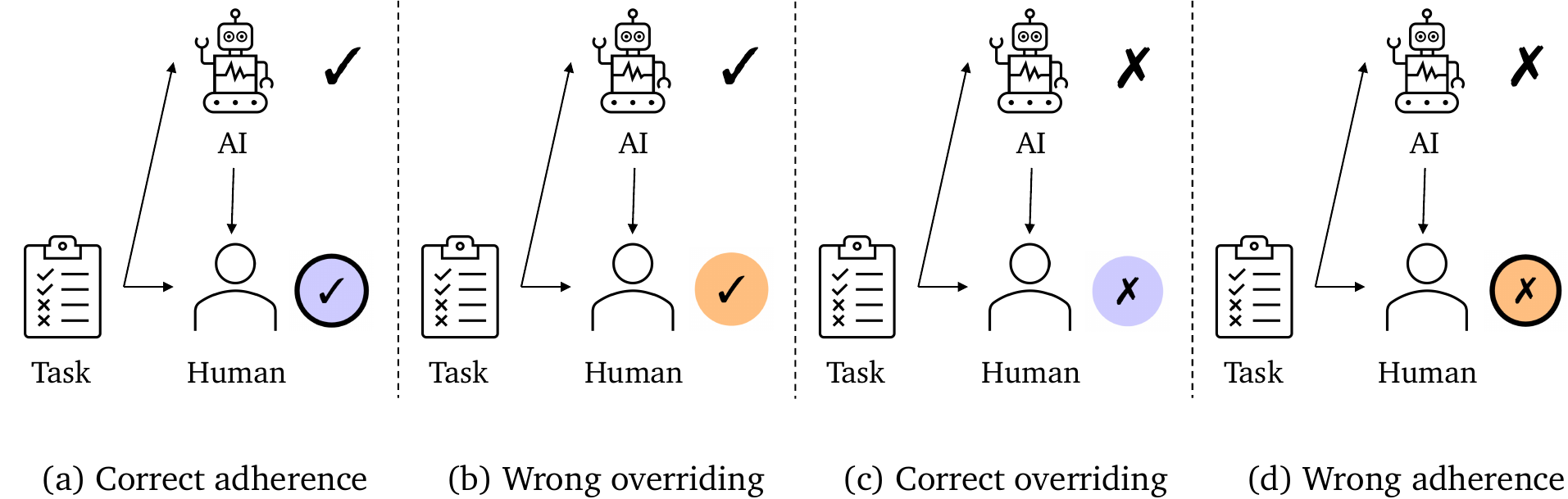}
    \caption{We consider \emph{concurrent} AI-assisted decision-making where a human-in-the-loop receives an AI recommendation that can either be correct (\ding{51}) or wrong (\ding{55}). The human can either adhere to (bordered circle) or override (no border) the AI recommendation. When the human adheres to a correct or overrides a wrong AI recommendation, the final decision will be correct (cases (a) and (c)); in the remaining cases, it will be wrong (cases (b) and (d)). The correctness of the final decision is indicated by either blue (correct) or orange (wrong) shading.}
    \label{fig:reliance_overview}
\end{figure}

Decision-making increasingly leverages support from artificial intelligence (AI)-based systems with the goal of making better and more efficient decisions.
Especially in high-stakes domains, such as lending, hiring, or healthcare, researchers and policymakers have often advocated for installing a human-in-the-loop as the ``last line of defense against AI failures''~\shortcite{passi2022overreliance}.
This approach assumes that humans are capable of rectifying such AI failures when they occur.
In AI-assisted decision-making, typically, an AI system generates an initial decision recommendation, which the human-in-the-loop may either adhere to or override (see \Cref{fig:reliance_overview}).
In order to complement the AI system, the human needs to adhere to AI recommendations if and only if these recommendations are correct, and override them otherwise.
Empirical studies have shown, however, that humans are often not able to achieve this type of \textit{appropriate reliance}\footnote{We use \emph{reliance} as an umbrella term for humans' behavior of adhering to or overriding AI recommendations~\shortcite{lai2023towards}.}~\shortcite{fok2023search}.
Instead, we often observe that they over- or under-rely on AI recommendations, indicating a deficiency to distinguish correct from wrong AI advice.
Even the introduction of additional means of decision support (e.g., explanations) has rarely produced the expected benefits in terms of human-AI complementarity~\shortcite{schemmer2022meta,schoeffer2024explanations}.
A key concern is that root cause analyses often struggle due to a limited understanding of $(i)$ how interventions influence human reliance on AI advice, and $(ii)$ how reliance relates to relevant metrics of decision quality.

In this work, we conceptually study the interplay of AI reliance and decision quality, and we highlight the importance of disentangling them in studies on AI-assisted decision-making.\footnote{This article extends the HHAI2023 conference paper by the authors~\shortcite{schoeffer2023interdependence}. It includes additional empirical analyses, a software implementation, and relevant extensions of our framework.}
We propose a framework that differentiates between reliance \textit{quantity} and \textit{quality}, enabling us to examine how each---both independently and in combination---affects decision quality metrics, particularly \textit{accuracy}.
Additionally, we characterize the conditions under which human-AI complementarity occurs, where human involvement improves decision quality compared to scenarios where the AI system operates autonomously.
Our framework is crafted to assist researchers and system designers in interpreting empirical findings, particularly the effects of interventions in AI-assisted decision-making.
To support this, we also offer an open-source tool built on our framework, which is available at \url{https://github.com/jhnnsjkbk/accuracy-reliance}.

This article is structured as follows. First, we provide relevant background and discuss related work in \Cref{sec:background}.
After that, in \Cref{sec:the_interdependence}, we formalize the relationship between AI reliance and decision-making accuracy, and we propose a visual framework to illustrate this relationship.
In \Cref{sec:interventions}, we show how our framework enables a meaningful interpretation of intervention effects. Crucially, we show that interventions can influence decision quality through significantly different effects on human reliance.
Without disentangling the effects on reliance and decision quality as we propose, two interventions may appear equally effective, when in reality, one leads to \textit{more} frequent overriding of AI recommendations, while the other results in \textit{fewer} overrides.
We discuss relevant extensions of our framework in \Cref{sec:extensions}, and we conclude by discussing implications, limitations, and avenues for future work in \Cref{sec:conclusion}.

\section{Background and Related Work}\label{sec:background}

In this section, we provide important background and discuss related work with respect to AI-assisted decision-making, reliance and trust, as well as relevant measures of decision quality and the effectiveness of interventions.

\subsection{AI-Assisted Decision-Making}
The utilization of AI-based systems in decision-making has been embraced across a multitude of critical domains~\shortcite{lai2023towards}.
In hiring, recent studies show that more than 55\% of human resource leaders in the United States use predictive algorithms to assist hiring activities~\shortcite{reicin2021ai}.
The underlying motives of adopting AI systems for assisting decision-making are diverse, including cost-cutting, improving decision quality, and enabling more robust and objective decisions~\shortcite{harris2005automated,kuncel2014hiring,lepri2018fair}.

The degree of AI integration in decision-making processes may vary depending on the specific context.
While many tasks may be well-suited for full automation through AI systems, others call for greater human oversight~\shortcite{de2020case}.
Particularly in high-stakes domains, AI systems often serve as decision support tools that aid human experts, who retain discretionary power to override AI recommendations and ultimately bear responsibility for making final decisions.
We refer to these human experts as the \textit{human-in-the-loop}.
For instance, in healthcare, AI systems can play a vital role in assisting clinicians with diagnoses or prognoses.
Subsequently, the clinicians can utilize these insights to determine the most appropriate course of treatment~\shortcite{leibig2022combining}.
Similarly, in the domain of criminal justice, judges might rely on AI-based risk assessment tools when determining bail~\shortcite{lima2021human}.

\begin{mdframed}
\begin{definition}[Human-in-the-loop]
    In the context of AI-assisted decision-making, \emph{human-in-the-loop} refers to a human decision-maker that retains discretionary power to override initially generated AI recommendations.
\end{definition}
\end{mdframed}

\begin{figure}[t]
    \centering
    \includegraphics[width=\textwidth]{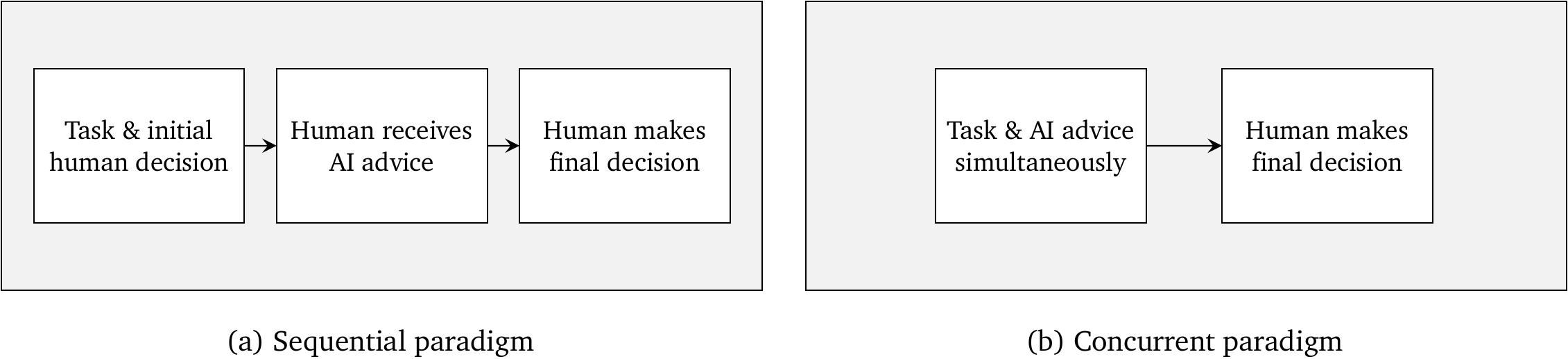}
    \caption{Two paradigms in AI-assisted decision-making: (a)~sequential and (b)~concurrent, taken from \shortciteA{tejeda2022ai}. The focus of this work is on \textit{concurrent} AI-assisted decision-making.}
    \label{fig:human-ai}
\end{figure}

In human-in-the-loop settings, an AI system typically generates an initial decision recommendation, which the human may either adhere to or override (see \Cref{fig:reliance_overview}).
In the taxonomy of \shortciteA{tejeda2022ai}, this corresponds to \emph{concurrent AI assistance}, where the human-in-the-loop does \emph{not} independently make a decision before AI assistance is provided (see \Cref{fig:human-ai}).
Prior work has also studied decision-making setups where a human makes an initial decision without AI support and then a final decision after receiving AI advice.
This \textit{sequential} paradigm has been empirically studied by \shortciteA{schemmer2023appropriate}, \shortciteA{dunning2023humans}, and \shortciteA{van2013framework}, among others.
One supposed advantage of the sequential setup is its ability to isolate the effect of AI advice on human decision-making by allowing researchers to observe whether and how the presence of such advice may change the initial human guess.
However, as pointed out by \shortciteA{tejeda2022ai}, many decision-making scenarios in the real world do not include independent human-only decisions prior to AI involvement.
Moreover, the sequential paradigm bears the risk of humans anchoring their second decision in their initial guess~\shortcite{echterhoff2022ai}.
For these reasons, we build on prior work~\shortcite{schoeffer2023interdependence} and focus on the \textit{concurrent} paradigm, stressing that our work complements related work on sequential decision-making.
Finally, we focus on binary decision-making as it is the most commonly studied type of decisions~\shortcite{lai2023towards}, e.g., in the realms of lending (loan/denial), hiring (offer/rejection), or recidivism prediction (release/detention).

\subsection{Reliance and Trust in AI-Assisted Decision-Making}

In concurrent AI-assisted decision-making, the human-in-the-loop is provided with an initial AI recommendation and may then leverage their discretionary power to either adhere to or override said recommendation.
In line with prior work, we call this behavior \textit{reliance}~\shortcite{eckhardt2024survey,Vereschak2021}.
\begin{mdframed}
\begin{definition}[AI reliance]
    In the context of AI-assisted decision-making, \emph{AI reliance} refers to the behavior of the human-in-the-loop that manifests in adhering to or overriding AI recommendations.
\end{definition}
\end{mdframed}
Measuring and manipulating the human reliance on AI recommendations has become a central pillar of research on AI-assisted decision-making \shortcite{fok2023search,schemmer2023appropriate,lai2023towards}. 
This is especially important as both humans and AI systems are commonly thought of as imperfect ``decision-makers'' with individual strengths and weaknesses \shortcite{kamar2016directions}. 
For humans that are assisted by AI, it is therefore important to be able to identify in which cases an AI recommendation is correct vs. wrong. 
This ability to detect AI errors has also been studied through the lens of signal detection theory, which describes how people distinguish signal from noise under uncertainty~\shortcite{langer2024effective,hautus2021detection}.
In the context of AI-assisted decision-making, the reliance behavior of overriding wrong AI recommendations and adhering to correct ones is typically referred to as \textit{appropriate reliance}~\shortcite{schemmer2023appropriate}.

Empirically, however, prior work has shown that humans are often unable to achieve appropriate reliance, even in the presence of different types of decision aids, such as explanations~\shortcite{fok2023search,schemmer2022meta}.
Instead, we often observe reliance behavior of the following types: $(i)$ overriding \textit{too many} AI recommendations (\textit{under-reliance}), or $(ii)$ overriding \textit{too few} AI recommendations (\textit{over-reliance}).
For instance, \shortciteA{dunning2023humans} find that humans tend to over-rely on highly skilled AI agents but under-rely when the AI agent is lower skilled.
Prior research, including \shortciteA{buccinca2021trust} and \shortciteA{kim2023algorithms}, has identified different scenarios in which under- or over-reliance results in reduced decision quality.
At the same time, prior work often lacks important nuance in discussing the effects of reliance on decision quality, such as when over-reliance vs. under-reliance might be preferable, and how instances of inappropriate reliance may still justify the presence of the human-in-the-loop.
Our work addresses this gap and complements recent studies by \shortciteA{fok2023search} and \shortciteA{guo2024decision}.

Also related to our work is prior work on trust calibration---i.e., achieving levels of \textit{appropriate trust} in AI systems, as opposed to over- or under-trust~\shortcite{de2020towards,sheridan2019extending}.
Many prior studies have treated \textit{reliance} and \textit{trust} interchangeably, sometimes calling reliance a ``behavioral trust measure''~\shortcite{papenmeier2022s}.
However, we stress that reliance and trust are different constructs:
reliance is commonly defined as the behavior of adhering to or overriding AI recommendations~\shortcite{eckhardt2024survey}, whereas trust is a subjective attitude regarding the whole system, which builds up and develops over time~\shortcite{parasuraman1997humans,rempel1985trust,schaefer2016meta,glikson2020human}.
It has been argued that trust may impact
reliance~\shortcite{dzindolet2003role,lee2004trust,shin2019role}, but trust is neither necessary nor sufficient for reliance when other factors, such as time constraints,
perceived risk, or self-confidence, impact decision-making~\shortcite{de2020case,lee2004trust,riley2018operator}.
In this work, we propose a taxonomy specifically for reliance behavior to facilitate a meaningful interpretation of empirical results in AI-assisted decision-making.

\subsection{Measuring Decision Quality}
Measuring the quality\footnote{We understand \textit{decision quality} as an umbrella term that subsumes any metrics that are a function of correct and wrong decisions. This excludes measures like efficiency, among others.} of AI-assisted decisions can be thought of as a two-stage process, in which the final decision quality depends on $(i)$ the initial quality of AI recommendations, and $(ii)$ the degree to which the human-in-the-loop can or cannot correct any AI mistakes.
In other words, the quality of AI-assisted decisions immediately depends on the degree to which the human adheres to or overrides AI recommendations, and \textit{how} they do so~\shortcite{jakubik2023empirical}.
For instance, even if the AI system issues 100\% correct recommendations, the final decision quality will decrease unless the human adheres to \textit{all} recommendations.
On the other hand, a relatively low-performance AI recommender may still result in high-quality decisions if the human-in-the-loop overrides all AI mistakes.
A relevant line of prior work has studied how different degrees of reliability of automation support affect human decision-making~\shortcite{dunning2023humans,rovira2007effects,wickens2007benefits,de2011adaptive}.
For instance, \shortciteA{dunning2023humans} show that humans tend to quickly over-rely on high-performance automation but under-rely if reliability is low.
This shows that human reliance behavior may itself be a function of the initial quality of AI recommendations, stressing the complexity of the interplay between AI reliance and decision quality, which motivates our work.

The \textit{accuracy} metric represents a key quality measure of AI-assisted decisions.
\textit{Accuracy} is defined as follows:
\begin{equation*}
    Accuracy = \frac{\# correct\ decisions}{\# decisions}.
\end{equation*}
Accuracy is frequently used for measuring the quality of AI-assisted decision-making~\shortcite{lai2023towards} and evaluating the effectiveness of interventions (e.g., explanations) for decision support~\shortcite{zhang2020effect,lai2020chicago,cabrera2023improving,lai2023towards}.
Note that we can also compute the accuracy of AI recommendations (``AI accuracy'').
Regarding the accuracy of decisions, prior works have studied whether humans can complement AI systems, i.e., improve the final decision-making accuracy over the AI accuracy (see \shortciteA{lai2023towards} for an overview).
In fact, a common motivation for providing humans with discretionary power is that they should be able to complement an AI system by overriding its wrong recommendations~\shortcite{de2020case}.
When the human-in-the-loop achieves that, we refer to this as \textit{complementarity}.\footnote{Other works have also referred to this as the \textit{team performance} of human and AI being greater than the performance of the individual agents alone~\shortcite{de2011adaptive,ijcai2022p344}.}
\begin{mdframed}
\begin{definition}[Complementarity]
    We call \emph{complementarity} the state when the human-in-the-loop relies on AI recommendations in a way that the final decision quality is higher than without human involvement.
\end{definition}
\end{mdframed}

In the context of the interplay of reliance and accuracy, this means that when the human-in-the-loop adheres to all correct and overrides all wrong AI recommendations, we achieve a final decision-making accuracy of 100\%.
However, as stated earlier, empirical evidence shows that achieving appropriate reliance is difficult in practice~\shortcite{fok2023search,schemmer2022meta}.
In cases where humans do \textit{not} appropriately rely on AI recommendations, the relationship between reliance behavior and accuracy is less obvious.
Crucially, there may be cases where \textit{inappropriate} reliance may still lead to complementarity.
These cases are not sufficiently well understood in the literature.

\subsection{Measuring Effectiveness of Interventions}
An emerging stream of research is studying decision support interventions that aim at facilitating complementarity~\shortcite{schemmer2023appropriate,cabitza2023ai}.
A prominent type of such interventions are explanations of various types~\shortcite{gunning2019darpa,arrieta2020explainable,adadi2018peeking,gilpin2018explaining}.
However, prior work has often focused on assessing the effectiveness of interventions only with respect to \textit{either} decision-making accuracy \shortcite{lai2023towards,kim2023algorithms} \textit{or} the human reliance on AI recommendations \shortcite{buccinca2021trust,lu2021human}.

Regarding \textit{accuracy}, it is commonly claimed that explanations are an enabler for better decisions~\shortcite{arrieta2020explainable,dodge2019explaining,gilpin2018explaining,rader2018explanations}.
A meta-study by \shortciteA{schemmer2022meta}, however, suggests that explanations in most empirical studies do not yield any significant benefits regarding decision-making accuracy; e.g., in the studies of \shortciteA{alufaisan2021does,green2019principles,liu2021understanding}, and \shortciteA{zhang2020effect}.
On the other hand, \shortciteA{lai2019human} find that explanations may enhance decision-making accuracy for the
case of deception detection.
Regarding \textit{reliance}, several studies find that explanations may increase \textit{or} decrease adherence of humans to AI recommendations regardless of their correctness (i.e., over- or under-reliance), such as the works by \shortciteA{bansal2021does,poursabzi2021manipulating,van2021evaluating}, and \shortciteA{schoeffer2024explanations}.

In this work, we argue that measuring \textit{either} accuracy \textit{or} reliance alone may provide an incomplete view when assessing AI-assisted decision-making generally and the effectiveness of interventions specifically.
More concretely, our proposed framework shows that $(i)$ focusing on accuracy alone is insufficient because an intervention may lead to an increase in decision quality by solely making a human adhere to more AI recommendations, and $(ii)$ focusing on reliance alone is insufficient because both over- and under-reliance may under certain circumstances still lead to improvements in accuracy compared to the baseline without human involvement.

\section{The Interdependence of Reliance Behavior and Accuracy}\label{sec:the_interdependence}

We consider binary decision-making tasks of $n\in\mathbf{N}$ instances with $n$ AI recommendations.
Let $Acc_{AI} \in (50\%,100\%]$ be the AI accuracy\footnote{Note that we only consider cases where the AI performs strictly better than chance.}, and $\mathcal{A} \in [0\%,100\%]$ the degree of human adherence to AI recommendations---e.g., $\mathcal{A}=70\%$ when the human-in-the-loop adheres to 70\% of AI recommendations.
As introduced in \Cref{fig:reliance_overview}, adherence can be correct ($\mathcal{A}_{correct}$) or wrong ($\mathcal{A}_{wrong}$).
We measure the correct and wrong adherence in percent, and we have $\mathcal{A}=\mathcal{A}_{correct}+\mathcal{A}_{wrong}$.
Similarly, we call the percentage of overrides $\mathcal{O} \in [0\%,100\%]$ (correct: $\mathcal{O}_{correct}$ or wrong: $\mathcal{O}_{wrong}$), and we have $\mathcal{O}=\mathcal{O}_{correct}+\mathcal{O}_{wrong}$.
We call the final decision-making accuracy, i.e., after human reliance behavior, $Acc_{final}$.
While in practice humans can only adhere to or override a finite number of AI recommendations, we often consider $n\rightarrow\infty$ for our theoretical considerations, so as to avoid rounding.
We summarize all notation in \Cref{tab:notation}.
Note that by definition we have:
\begin{equation}\label{eq:def_relationships}
    \begin{aligned}
        \mathcal{A}+\mathcal{O} &=\mathcal{A}_{correct}+\mathcal{A}_{wrong}+\mathcal{O}_{correct}+\mathcal{O}_{wrong}=100\% \\
        Acc_{AI} &=\mathcal{A}_{correct}+\mathcal{O}_{wrong} \\
        Acc_{final} &=\mathcal{A}_{correct}+\mathcal{O}_{correct}.
    \end{aligned}
\end{equation}
 
\begin{table}[t]
    \centering
    \begin{tabular}{ll}
    \toprule
        Symbol & Description \\
        \midrule
        $Acc_{AI}$ & AI accuracy in \% \\
        $Acc_{final}$ & (Final) decision-making accuracy in \%, after human reliance \\ 
        $\mathcal{A}$ & Degree of human adherence to AI recommendations in \% \\
        $\mathcal{A}_{correct}$ & Correct adherence in \%: \textbf{adhering} to \textbf{correct} AI recommendation \\
        $\mathcal{A}_{wrong}$ & Wrong adherence in \%: \textbf{adhering} to \textbf{wrong} AI recommendation \\
        $\mathcal{O}$ & Degree of human overriding of AI recommendations in \% \\
        $\mathcal{O}_{correct}$ & Correct overriding in \%: \textbf{overriding} of \textbf{wrong} AI recommendation \\
        $\mathcal{O}_{wrong}$ & Wrong overriding in \%: \textbf{overriding} of \textbf{correct} AI recommendation \\
        $n\in\mathbf{N}$ & Number of AI recommendations \\
        $Q \in [0,1]$ & Quality of human reliance \\
        \ding{51} & Correct AI recommendation \\
        \ding{55} & Wrong AI recommendation \\
    \bottomrule
    \end{tabular}
    \caption{Summary of important notation.}
    \label{tab:notation}
\end{table}

\subsection{Motivational Example}\label{sec:motivational}

Consider the following motivational example: we have a task that consists of making $n=10$ binary decisions.
The AI system that is used for providing decision recommendations to the human has an accuracy of $Acc_{AI}=70\%$; i.e., 7 out of 10 recommendations are correct (\ding{51}) and 3 are wrong (\ding{55}).
Now, when the human-in-the-loop adheres to all AI recommendations ($\mathcal{A}=100\%$), this leads to a decision-making accuracy of $Acc_{final}=70\%$, equal to the AI accuracy.
In terms of reliance behavior, this implies that the human-in-the-loop correctly adheres to 7 correct AI recommendations ($\mathcal{A}_{correct}=70\%$), and wrongly adheres to the remaining 3 recommendations ($\mathcal{A}_{wrong}=30\%$).
In the other extreme case, where the human-in-the-loop overrides all AI recommendations ($\mathcal{O}=100\%$), the resulting decision-making accuracy will be $100\%-70\%=30\%$, where the human correctly overrides 3 wrong AI recommendations ($\mathcal{O}_{correct}=30\%$), and wrongly overrides 7 correct AI recommendations ($\mathcal{O}_{wrong}=70\%$).

If the human reliance behavior is mixed, i.e., when the human-in-the-loop adheres to some AI recommendations and overrides others, decision-making accuracy will depend on how well they can distinguish cases where the AI is correct from cases where it is wrong.
To make this clear, consider the same AI system as above with an accuracy of 70\%, and a human-in-the-loop that adheres to 7 out of 10 of its recommendations ($\mathcal{A}=70\%$).
This is illustrated in \Cref{fig:sys_acc_examples}.
\begin{figure}[t]
     \centering
        \begin{subfigure}[a]{\textwidth}
        \centering
        \begin{tikzpicture}[]
        \node[circle,draw, ultra thick, text=black,fill=blue!20] (c) at (0,0){\ding{51}};
        \node[circle,draw, ultra thick, text=black,fill=blue!20] (c) at (1,0){\ding{51}};
        \node[circle,draw, ultra thick, text=black,fill=blue!20] (c) at (2,0){\ding{51}};
        \node[circle,draw, ultra thick, text=black,fill=blue!20] (c) at (3,0){\ding{51}};
        \node[circle,draw, ultra thick, text=black,fill=blue!20] (c) at (4,0){\ding{51}};
        \node[circle,draw, ultra thick, text=black,fill=blue!20] (c) at (5,0){\ding{51}};
        \node[circle,draw, ultra thick, text=black,fill=blue!20] (c) at (6,0){\ding{51}};
        \draw[dashed] (6.5,-0.5) -- (6.5,0.5);
        \node[circle,text=black,fill=blue!20] (c) at (7,0){\ding{55}};
        \node[circle,text=black,fill=blue!20] (c) at (8,0){\ding{55}};
        \node[circle,text=black,fill=blue!20] (c) at (9,0){\ding{55}};
        \end{tikzpicture}
        \caption{$Acc_{final}=100\%$}
        \label{fig:sys_acc_100}
     \end{subfigure}
     \begin{subfigure}[b]{\textwidth}
          \centering
           \begin{tikzpicture}[]
            \node[circle,text=black,fill=orange!50] (c) at (0,0){\ding{51}};
            \node[circle,draw, ultra thick, text=black,fill=blue!20] (c) at (1,0){\ding{51}};
            \node[circle,draw, ultra thick, text=black,fill=blue!20] (c) at (2,0){\ding{51}};
            \node[circle,draw, ultra thick, text=black,fill=blue!20] (c) at (3,0){\ding{51}};
            \node[circle,draw, ultra thick, text=black,fill=blue!20] (c) at (4,0){\ding{51}};
            \node[circle,draw, ultra thick, text=black,fill=blue!20] (c) at (5,0){\ding{51}};
            \node[circle,draw, ultra thick, text=black,fill=blue!20] (c) at (6,0){\ding{51}};
            \draw[dashed] (6.5,-0.5) -- (6.5,0.5);
            \node[circle,draw, ultra thick, text=black,fill=orange!50] (c) at (7,0){\ding{55}};
            \node[circle,text=black,fill=blue!20] (c) at (8,0){\ding{55}};
            \node[circle, text=black,fill=blue!20] (c) at (9,0){\ding{55}};
            \end{tikzpicture}
        \caption{$Acc_{final}=80\%$}
        \label{fig:sys_acc_80}
     \end{subfigure}
     \begin{subfigure}[c]{\textwidth}
     \centering
        \begin{tikzpicture}[]
            \node[circle,text=black,fill=orange!50] (c) at (0,0){\ding{51}};
            \node[circle,text=black,fill=orange!50] (c) at (1,0){\ding{51}};
            \node[circle,draw, ultra thick,text=black,fill=blue!20] (c) at (2,0){\ding{51}};
            \node[circle,draw, ultra thick,text=black,fill=blue!20] (c) at (3,0){\ding{51}};
            \node[circle,draw, ultra thick,text=black,fill=blue!20] (c) at (4,0){\ding{51}};
            \node[circle,draw, ultra thick,text=black,fill=blue!20] (c) at (5,0){\ding{51}};
            \node[circle,draw, ultra thick,text=black,fill=blue!20] (c) at (6,0){\ding{51}};
            \draw[dashed] (6.5,-0.5) -- (6.5,0.5);
            \node[circle,draw, ultra thick,text=black,fill=orange!50] (c) at (7,0){\ding{55}};
            \node[circle,draw, ultra thick,text=black,fill=orange!50] (c) at (8,0){\ding{55}};
            \node[circle, text=black,fill=blue!20] (c) at (9,0){\ding{55}};
        \end{tikzpicture}
        \caption{$Acc_{final}=60\%$}
        \label{fig:sys_acc_60}
     \end{subfigure}
     \vspace{0.5cm}
     \begin{subfigure}[d]{\textwidth}
     \centering
        \begin{tikzpicture}[]
             \node[circle,text=black,fill=orange!50] (c) at (0,0){\ding{51}};
            \node[circle,text=black,fill=orange!50] (c) at (1,0){\ding{51}};
            \node[circle,text=black,fill=orange!50] (c) at (2,0){\ding{51}};
            \node[circle,draw, ultra thick,text=black,fill=blue!20] (c) at (3,0){\ding{51}};
            \node[circle,draw, ultra thick,text=black,fill=blue!20] (c) at (4,0){\ding{51}};
            \node[circle,draw, ultra thick,text=black,fill=blue!20] (c) at (5,0){\ding{51}};
            \node[circle,draw, ultra thick,text=black,fill=blue!20] (c) at (6,0){\ding{51}};
            \draw[dashed] (6.5,-0.5) -- (6.5,0.5);
            \node[circle,draw, ultra thick,text=black,fill=orange!50] (c) at (7,0){\ding{55}};
            \node[circle,draw, ultra thick,text=black,fill=orange!50] (c) at (8,0){\ding{55}};
            \node[circle,draw, ultra thick, text=black,fill=orange!50] (c) at (9,0){\ding{55}};
        \end{tikzpicture}
        \caption{$Acc_{final}=40\%$}
        \label{fig:sys_acc_40}
     \end{subfigure}
        \begin{tikzpicture}[
                node distance={20mm},
                thick,
                main/.style = {draw, circle, inner sep=2pt, minimum size=8mm},
                box/.style = {draw,black,inner sep=10pt,rounded corners=5pt}] 
            
            \node[circle,draw, ultra thick,text=black,fill=blue!20] (1) at (0,0){\ding{51}};
            \node[text=black,anchor=west] (c1) at (0.4,0){Correct adherence};
            \node[circle, text=black,fill=orange!50] (2) at (0,-1){\ding{51}};
            \node[text=black,anchor=west] (c2) at (0.4,-1){Wrong override};
            
            \node[circle, draw, ultra thick, text=black,fill=orange!50] (3) at (4.5,0){\ding{55}};
            \node[text=black, anchor=west] (c3) at (4.9,0){Wrong adherence};
            \node[circle, text=black,fill=blue!20] (4) at (4.5,-1){\ding{55}};
            \node[text=black, anchor=west] (c4) at (4.9,-1){Correct override};

            \node[box,fit=(1)(2)(c3)(c4)]{};
        \end{tikzpicture}
     \caption{Possible scenarios of reliance behavior and associated decision-making accuracy, given an AI accuracy of $Acc_{AI}=70\%$ and an adherence level of $\mathcal{A}=70\%$. Correct AI recommendations (\ding{51}) and wrong AI recommendation (\ding{55}) are separated by a dashed line.}
     \label{fig:sys_acc_examples}
\end{figure}
If the human-in-the-loop is able to perfectly distinguish between correct and wrong AI recommendations, they will adhere to all 7 correct AI recommendations ($\mathcal{A}_{correct}=70\%=\mathcal{A}$) and override the 3 wrong ones ($\mathcal{O}_{correct}=30\%=\mathcal{O}$).
The resulting decision-making accuracy would then be $Acc_{final}=100\%$ (case (a) in \Cref{fig:sys_acc_examples}).
In this case, the human-in-the-loop is able to perfectly complement the AI by correcting for its mistakes.
Cases~(b)--(d) in \Cref{fig:sys_acc_examples} show situations where the human-in-the-loop still adheres to 70\% of AI recommendations but their ability to override wrong AI recommendations decreases.
For instance, consider case (d), where the human-in-the-loop does not perform any correct overrides ($\mathcal{O}_{correct}=0$).
If the degree of human adherence to AI recommendations is fixed at 70\%, this is, in fact, the worst possible reliance behavior with respect to accuracy, resulting in a decision-making accuracy of $Acc_{final}=40\%$.

From \Cref{fig:sys_acc_examples}, we can also infer that if the human-in-the-loop overrides \textit{more} than 3 AI recommendations, at least one of these overrides must be wrong (i.e., the human would override a correct AI recommendation), meaning that a decision-making accuracy of 100\% would no longer be possible.
We may think of such a reliance behavior as \textit{under-reliance}.
Similarly, when the human overrides \textit{less} than 3 AI recommendations, there must be at least one instance of wrong adherence.
This might be referred to as \textit{over-reliance}.
Generally, we may think of under-reliance as a behavior where $\mathcal{A}<Acc_{AI}$, and over-reliance as $\mathcal{A}>Acc_{AI}$.
Note that there exists other work that has been thinking of these terms with respect to behavior at the level of individual decisions~\shortcite{schemmer2023appropriate,vasconcelos2022explanations}.

\subsection{The General Case}
Generally, any degree of adherence to AI recommendations is associated with a range of possible decision-making accuracy, based on how well the human-in-the-loop can override the AI recommendations when they are wrong and adhere to them when they are correct.
In \Cref{fig:sys_acc_examples}, this range would be $Acc_{final}\in\{40\%,60\%,80\%,100\%\}$ for $n=10$, a given AI accuracy of $Acc_{AI}=70\%$, and a degree of adherence to AI recommendations of $\mathcal{A}=70\%$.
As mentioned earlier, we generally consider $n\rightarrow\infty$, in which case the possible values in this range become continuous.
We state the following proposition on the attainable decision-making accuracy as a function of the AI accuracy as well as the degree of human adherence to AI recommendations.
\begin{mdframed}
\begin{proposition}\label{prop:acc_final}
    For $n\rightarrow\infty$, a given AI accuracy $Acc_{AI}$, and a degree of adherence to AI recommendations, $\mathcal{A}$, the range of attainable decision-making accuracy $Acc_{final}$ is
    \[
    \small
    Acc_{final} \in 
    \begin{dcases*}
        [100\%-Acc_{AI}-\mathcal{A},100\%-Acc_{AI}+\mathcal{A}] & if\ \ $0\leq \mathcal{A} \leq 100\%-Acc_{AI}$ \\
        [-100\%+Acc_{AI}+\mathcal{A},100\%-Acc_{AI}+\mathcal{A}] & if\ \ $100\%-Acc_{AI} < \mathcal{A} \leq Acc_{AI}$ \\
        [-100\%+Acc_{AI}+\mathcal{A},100\%+Acc_{AI}-\mathcal{A}] & if\ \ $Acc_{AI} < \mathcal{A} \leq 100\%$.
    \end{dcases*}
    \]
\end{proposition}
\end{mdframed}
The maximum of this accuracy range will be attained whenever the human-in-the-loop maximizes correct adherence and correct overrides given a degree of adherence $\mathcal{A}$, since $Acc_{final}=\mathcal{A}_{correct}+\mathcal{O}_{correct}$.
Hence, in the ideal case, we would have $\mathcal{A}_{correct}+\mathcal{O}_{correct}=100\%$, which immediately implies that $\mathcal{A}_{wrong}=\mathcal{O}_{wrong}=0\%$.
This would be case (a) in \Cref{fig:sys_acc_examples}.
However, as we can see in \Cref{prop:acc_final}, this is only possible when $\mathcal{A}=\mathcal{A}_{correct}=Acc_{AI}$, meaning that the human must adhere to AI recommendations if and only if they are correct, and override otherwise.
In other words, to achieve a decision-making accuracy of $Acc_{final}=100\%$, we need two conditions:
\begin{itemize}
    \item[$(i)$] The general degree of adherence to AI recommendations, $\mathcal{A}$, is equal to the AI accuracy $Acc_{AI}$, i.e., $\mathcal{A}=Acc_{AI}$.
    \item[$(ii)$] The human-in-the-loop must be able to adhere to any correct AI recommendation and override any wrong one, i.e., $\mathcal{A}_{correct}=\mathcal{A}$ and $\mathcal{O}_{correct}=\mathcal{O}$.
\end{itemize}
However, in practice, it is likely that either $(i)$ or $(ii)$ are not satisfied and, hence, the decision-making accuracy is less than 100\%.
Even if $(i)$ is satisfied, like in \Cref{fig:sys_acc_examples}, we see in cases (b)--(d) that $Acc_{final}$ is negatively affected when humans adhere to wrong AI recommendations and override correct ones.

\subsection{A Visual Framework}\label{subsec:visual_framework}
\begin{figure}[ht]
    \centering
    \begin{minipage}[t]{0.48\textwidth}
        \centering
        \includegraphics[width=0.95\textwidth]{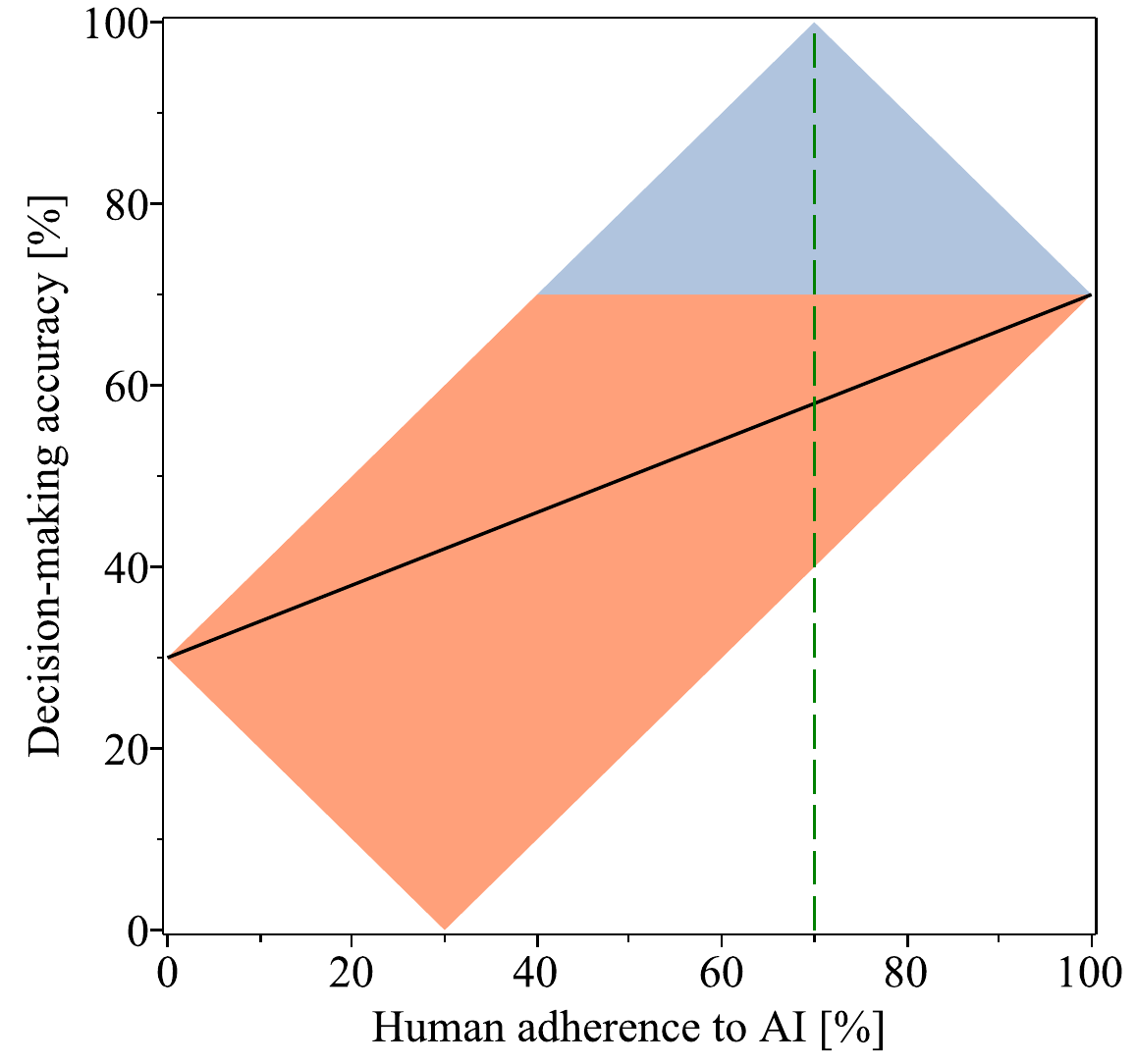}
        (a) $Acc_{AI}=70\%$
    \end{minipage}\hfill
    \begin{minipage}[t]{0.48\textwidth}
        \centering
        \includegraphics[width=0.95\textwidth]{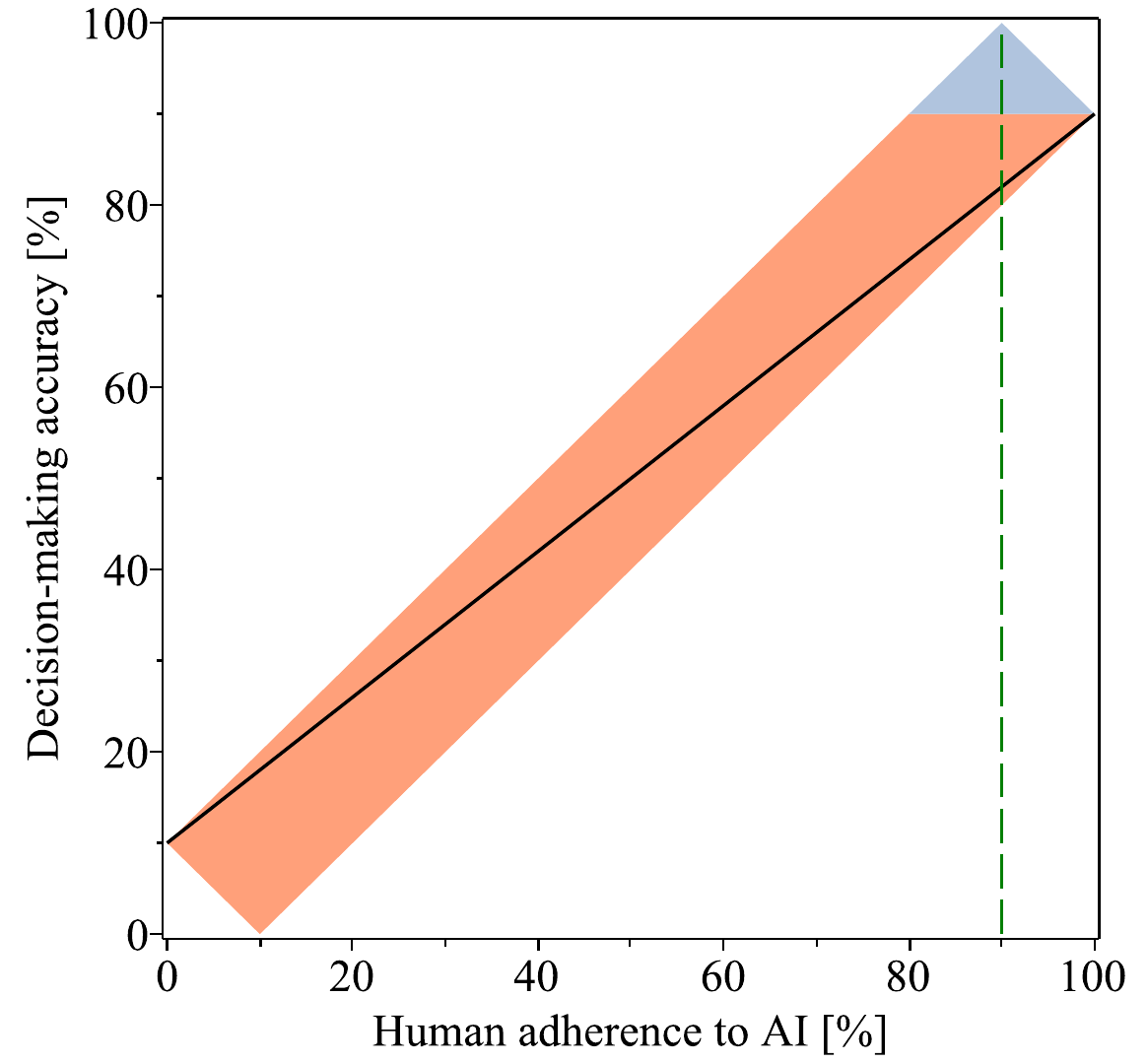}
        (b) $Acc_{AI}=90\%$
    \end{minipage}\hfill
\caption{The area of attainable decision-making accuracy for a given AI accuracy of (a)~70\% and (b)~90\%, and different levels of human adherence. The orange area indicates $Acc_{final}<Acc_{AI}$; blue indicates $Acc_{final}>Acc_{AI}$; the green dashed line indicates the level of adherence where $Acc_{final}=100\%$ is attainable; the black line indicates the expected value of $Acc_{final}$ when the human-in-the-loop cannot discern correct and wrong.}
\label{fig:framework_dev}
\end{figure}

To make the general relationship between reliance behavior and decision-making accuracy more tangible, we visualize \Cref{prop:acc_final} in \Cref{fig:framework_dev} for (a) $Acc_{AI}=70\%$ and (b) $Acc_{AI}=90\%$.
On the horizontal axes we have the human adherence to AI recommendations, $\mathcal{A}\in[0,100\%]$.
The vertical axes show the decision-making accuracy, $Acc_{final}\in[0,100\%]$.
The filled rectangular area in orange and blue combined constitutes the attainable decision-making accuracy for any given $\mathcal{A}$.
We distinguish orange and blue to highlight areas where the human-in-the-loop complements the AI (blue, $Acc_{final}>Acc_{AI}$) or impairs it (orange, $Acc_{final}<Acc_{AI}$) regarding accuracy.
The green dashed vertical line indicates the level of $\mathcal{A}=Acc_{AI}$, which corresponds to the degree of adherence where the maximum decision-making accuracy of 100\% can be attained, as discussed previously.
Note that as the AI accuracy increases (\Cref{fig:framework_dev} (a) $\rightarrow$ (b)), the colored area decreases; and for $Acc_{AI}=100\%$ it becomes a line, in which case \Cref{prop:acc_final} collapses into $Acc_{final}=\mathcal{A}$.
This means that the better the AI, the smaller the variance for different reliance behaviors. 

Contrasting the orange and blue areas, we immediately see that up to a certain level of adherence $\mathcal{A}$ there is no possibility to reach the blue area of complementarity.
We also see that the minimum level of $\mathcal{A}$ for which the human-in-the-loop may complement the AI increases as $Acc_{AI}$ increases ($\mathcal{A}=40\%$ in \Cref{fig:framework_dev} (a) $\rightarrow$ $\mathcal{A}=80\%$ in \Cref{fig:framework_dev} (b)).
Finally, when $\mathcal{A}\geq Acc_{AI}$, attaining a decision-making accuracy in the blue area is always \textit{possible}.
We characterize this in the following corollary:
\begin{mdframed}
\begin{corollary}\label{cor:under-rely}
    When the human-in-the-loop under-relies at a degree of $\mathcal{A}<2\cdot Acc_{AI}-100\%$, we will always have $Acc_{final}<Acc_{AI}$. When $\mathcal{A}>2\cdot Acc_{AI}-100\%$, achieving a decision-making accuracy greater than the AI accuracy, i.e., $Acc_{final}>Acc_{AI}$, is possible.
\end{corollary}
\end{mdframed}

From the visual framework, we can also formally infer that any $Acc_{final}\in (0,100\%)$ can be associated with different degrees of adherence $\mathcal{A}$.
In fact, due to the symmetric shape of the rectangle, when we think of $Acc_{final}$ as a function of $\mathcal{A}$, the inverse $\mathcal{A}(Acc_{final})$ would be identical to the function itself.
For instance, a decision-making accuracy of $Acc_{final}=70\%$ may correspond to any $\mathcal{A}\in[40\%,100\%]$ in \Cref{fig:framework_dev}~(a).
\begin{mdframed}
\begin{proposition}\label{prop:multiple_reliance}
    When $Acc_{final}(\mathcal{A})\in[u,v]$ for a given $\mathcal{A}$, we have $\mathcal{A}(Acc_{final}) \in [u,v]$.
\end{proposition}
\end{mdframed}
However, when fixing $Acc_{final}$ at 70\% in \Cref{fig:framework_dev}~(a), different levels of $\mathcal{A}$ correspond to different vertical positions within the rectangle: $\mathcal{A}=40\%$ corresponds to a position at the very northern border of the rectangle, whereas any $\mathcal{A}>40\%$ corresponds to a position on the horizontal line separating the orange and blue areas, which leaves room for accuracy improvements.
This means that a given decision-making accuracy can be achieved through strikingly different qualities of reliance.
We address this, as well as the role of the black separating lines in \Cref{fig:framework_dev}, in more detail in the following.

\subsection{Discerning Correct and Wrong AI Recommendations}\label{subsec:discerning}
While a horizontal movement in the framework constitutes a change in the \textit{quantity} of adherence to AI recommendations, this information alone does not capture the \textit{quality} of reliance---this information is captured in the vertical movements.
To make this more concrete, consider again a task with AI recommendations that are 70\% accurate.
When the human-in-the-loop has no ability to distinguish correct from wrong AI recommendations, the likelihood of adhering to or overriding a given AI recommendation is the same regardless of whether that recommendation is correct or wrong. 
Hence, at an adherence of $\mathcal{A}$, we would expect the human-in-the-loop to adhere to $\mathcal{A}\%$ of correct AI recommendations and $\mathcal{A}\%$ of wrong AI recommendations.
At $Acc_{AI}=70\%$, this implies that $\mathcal{A}\%$ of 70\% are correct adherences, $\mathcal{A}\%$ of 30\% are wrong adherences, $(100-\mathcal{A})\%$ of 70\% are wrong overrides, and $(100-\mathcal{A})\%$ of 30\% are correct overrides.
When we have $\mathcal{A}=70\%$, this would imply $\mathcal{A}_{correct}=49\%$, $\mathcal{A}_{wrong}=21\%$, $\mathcal{O}_{correct}=9\%$, and $\mathcal{O}_{wrong}=21\%$, with a decision-making accuracy of $\mathcal{A}_{correct}+\mathcal{O}_{correct}=58\%$.
This corresponds to the intersection of the black line with the dashed green vertical line in \Cref{fig:framework_dev} (a).
We generalize this in the following proposition.
\begin{mdframed}
\begin{proposition}\label{prop:discern}
    When the human-in-the-loop cannot discern correct and wrong AI recommendations, the expected decision-making accuracy is linearly increasing in $\mathcal{A}$ and given by 
    \begin{align*}
        Acc_{final} (\mathcal{A})&=\mathcal{A}\cdot Acc_{AI}+(100\%-\mathcal{A})\cdot (100\%-Acc_{AI}) \\
        &=(100\%-Acc_{AI}) + \underbrace{(2\cdot Acc_{AI}-100\%)}_{>0}\cdot \mathcal{A},
    \end{align*}
    for a given AI accuracy $Acc_{AI}$.
\end{proposition}
\end{mdframed}
Note that the relationship from \Cref{prop:discern} equates to the black lines in \Cref{fig:framework_dev}, which separate the respective rectangles in half.
We immediately see the following:
\begin{mdframed}
\begin{corollary}\label{cor:lower}
    When the human-in-the-loop cannot discern correct and wrong AI recommendations, the expected decision-making accuracy is always lower than or equal to the AI accuracy, i.e., $Acc_{final}\leq Acc_{AI}$.
\end{corollary}
\end{mdframed}

Having established the expected decision-making accuracy when the human-in-the-loop is not able to distinguish correct and wrong AI recommendations, we now turn to cases where they can---to different degrees.
Such reliance behavior corresponds to points in the framework that are situated \textit{above} the black line. 
While certainly less relevant in practice, we might also think of cases where a human adheres to and overrides AI recommendations worse than chance, which would correspond to points \textit{below} the black line.
Following up on \Cref{prop:acc_final}, we now examine three cases based on different adherence levels, and we characterize the reliance behavior that is associated with the maximum and minimum decision-making accuracy for given $\mathcal{A}$.

\paragraph{Case: $0\leq \mathcal{A} \leq 100\%-Acc_{AI}$}
Since we assume that $Acc_{AI}>50\%$, we have $\mathcal{A}<Acc_{AI}$ in this case.
When the degree of adherence to AI recommendations is strictly smaller than the AI accuracy, achieving a decision-making accuracy of $Acc_{final}=100\%$ is no longer possible.
This also implies that there must be at least one instance where the human-in-the-loop overrides a correct AI recommendation, i.e., $\mathcal{O}_{wrong}>0$. 
From \Cref{prop:acc_final} we also see that the \textbf{maximum} achievable decision-making accuracy in that case is $100\%-Acc_{AI}+\mathcal{A}$, which is achieved when $\mathcal{A}=\mathcal{A}_{correct}$.
Using the definition of $\mathcal{A}$ and relationships from \Cref{eq:def_relationships}, this directly implies that $\mathcal{A}_{wrong}=0$, $\mathcal{O}_{correct}=100\%-Acc_{AI}$, and $\mathcal{O}_{wrong}=Acc_{AI}-\mathcal{A}>0$.
The \textbf{minimum} achievable decision-making accuracy, on the other hand, is attained when adherence only happens to wrong AI recommendations, hence, $\mathcal{A}_{wrong}=\mathcal{A}$.
Similar to above, this implies that $\mathcal{A}_{correct}=0$, $\mathcal{O}_{wrong}=Acc_{AI}$, and $\mathcal{O}_{correct}=100\%-Acc_{AI}-\mathcal{A}$.

To illustrate this, let us reconsider the example from \Cref{fig:sys_acc_examples}, but with a degree of adherence to AI recommendations of $\mathcal{A}=20\%$.
The attainable decision-making accuracy in this case is, according to \Cref{prop:acc_final}, $Acc_{final} \in [10\%,50\%]$.
To achieve $Acc_{final}=50\%$, the human-in-the-loop would have to adhere to 2 correct AI recommendations ($\mathcal{A}_{correct}=20\%$) and 0 wrong AI recommendations ($\mathcal{A}_{wrong}=0$).
The remaining 8 AI recommendations, 5 of which are correct and 3 wrong, are overridden (i.e., $\mathcal{O}_{wrong}=50\%$ and $\mathcal{O}_{correct}=30\%$).
The minimum decision-making accuracy of 10\%, on the other hand, is attained when the human-in-the-loop only adheres to wrong AI recommendations (i.e., $\mathcal{A}_{wrong}=20\%$ and $\mathcal{A}_{correct}=0$).
The remaining AI recommendations, 7 correct and 1 wrong, are overridden, which implies $\mathcal{O}_{wrong}=70\%$ and $\mathcal{O}_{correct}=10\%$.
Overall, we conclude the following:
\begin{mdframed}
\begin{corollary}
    When $0\leq \mathcal{A} \leq 100\%-Acc_{AI}$, the decision-making accuracy is maximal when all adherence is to correct AI recommendations (i.e., $\mathcal{A}_{correct}=\mathcal{A}$), and it is minimal when all adherence is to wrong AI recommendations (i.e., $\mathcal{A}_{wrong}=\mathcal{A}$).
\end{corollary}
\end{mdframed}

\paragraph{Case: $100\%-Acc_{AI} < \mathcal{A} \leq Acc_{AI}$}
With the same argument as in the previous case, the \textbf{maximum} decision-making accuracy is attained when $\mathcal{A}_{correct}=\mathcal{A}$, which directly implies $\mathcal{A}_{wrong}=0$, $\mathcal{O}_{correct}=100\%-Acc_{AI}$, and $\mathcal{O}_{wrong}=Acc_{AI}-\mathcal{A}$.
As for the \textbf{minimum} decision-making accuracy, note that since $\mathcal{A} > 100\%-Acc_{AI}$, we must have $\mathcal{A}_{correct}>0$, i.e., the human-in-the-loop must be adhering to at least one correct AI recommendation.
The minimum accuracy is thus attained when the human-in-the-loop adheres to all wrong AI recommendations (plus at least one correct recommendation).
This implies that all overrides must be of correct AI recommendations, i.e., we have $\mathcal{O}_{wrong}=\mathcal{O}$, $\mathcal{O}_{correct}=0$, as well as $\mathcal{A}_{correct}=Acc_{AI}-\mathcal{O}>0$, and $\mathcal{A}_{wrong}=100\%-Acc_{AI}$.
\begin{mdframed}
\begin{corollary}
    When $100\%-Acc_{AI} < \mathcal{A} \leq Acc_{AI}$, the decision-making accuracy is maximal when all adherence is to correct AI recommendations (i.e., $\mathcal{A}_{correct}=\mathcal{A}$), and it is minimal when all overrides are of correct AI recommendations (i.e., $\mathcal{O}_{wrong}=\mathcal{O}$).
\end{corollary}
\end{mdframed}

\paragraph{Case: $Acc_{AI} < \mathcal{A} \leq 100\%$}
While in the previous two cases we had $\mathcal{A}\leq Acc_{AI}$, we now consider the case where the human-in-the-loop over-relies on the AI recommendations, meaning that there must be at least one case where they adhere to a wrong AI recommendation, i.e., $\mathcal{A}_{wrong}>0$.
The \textbf{maximum} decision-making accuracy will thus be attained when all overrides are correct, i.e., $\mathcal{O}_{correct}=\mathcal{O}$, which immediately implies $\mathcal{O}_{wrong}=0$, $\mathcal{A}_{correct}=Acc_{AI}$, and $\mathcal{A}_{wrong}=100\%-Acc_{AI}-\mathcal{O}>0$.
The \textbf{minimum} decision-making accuracy, on the other hand, will be attained when all overrides are wrong, similar to the previous case. Hence, we would also observe $\mathcal{O}_{wrong}=\mathcal{O}$, $\mathcal{O}_{correct}=0$, $\mathcal{A}_{correct}=Acc_{AI}-\mathcal{O}$, and $\mathcal{A}_{wrong}=100\%-Acc_{AI}$.
\begin{mdframed}
\begin{corollary}\label{cor:over-reliance}
    When $Acc_{AI} < \mathcal{A} \leq 100\%$, the decision-making accuracy is maximal when all overrides are of wrong AI recommendations (i.e., $\mathcal{O}_{correct}=\mathcal{O}$), and it is minimal when all overrides are of correct AI recommendations (i.e., $\mathcal{O}_{wrong}=\mathcal{O}$).
\end{corollary}
\end{mdframed}

\subsection{Measuring the Quality of Reliance for Given $\mathcal{A}$}\label{sec:quality-of-reliance}
In the previous subsection, we established the reliance behavior that is associated with the extreme cases of maximum and minimum decision-making accuracy for any given degree of adherence to AI recommendations.
Now, we develop a metric $Q\in[0,1]$ for the quality of reliance given $Acc_{AI}$, such that a value of $Q=1$ corresponds to the maximum attainable decision-making accuracy, and $Q=0$ to the minimum. 
First, we derive the following corollary from \Cref{prop:acc_final}:
\begin{mdframed}
\begin{corollary}\label{cor:width}
    The width $W$ of the range of attainable values of $Acc_{final}$ is:
    \[
    W (\mathcal{A}) = 
    \begin{dcases*}
        2\cdot\mathcal{A} & if\ \ $0\leq \mathcal{A} \leq 100\%-Acc_{AI}$ \\
        2\cdot(100\%-Acc_{AI}) & if\ \ $100\%-Acc_{AI} < \mathcal{A} \leq Acc_{AI}$ \\
        2\cdot(100\%-\mathcal{A}) & if\ \ $Acc_{AI} < \mathcal{A} \leq 100\%$.
    \end{dcases*}
    \]
\end{corollary}
\end{mdframed}
Geometrically, $W$ corresponds to the distance between the upper and lower vertical boundary of the rectangle (see, e.g., \Cref{fig:framework_dev}) for a fixed $\mathcal{A}$.
With that, we can define our metric $Q$ as follows:
\begin{equation}\label{eq:quality}
    Q(\mathcal{A},Acc_{final})\coloneqq 
    \begin{dcases*}
        \frac{Acc_{final}-(100\%-Acc_{AI}-\mathcal{A})}{W(\mathcal{A})} & \textit{if}\ \ $0\leq \mathcal{A} \leq 100\%-Acc_{AI}$ \\
        \frac{Acc_{final}+(100\%-Acc_{AI}-\mathcal{A})}{W(\mathcal{A})} & \textit{if}\ \ $100\%-Acc_{AI} < \mathcal{A}$.
    \end{dcases*}
\end{equation}
If $Acc_{AI}$ and $\mathcal{A}$ are fixed, maximizing the quality of reliance corresponds to maximizing $Acc_{final}=\mathcal{A}_{correct}+\mathcal{O}_{correct}$ given $\mathcal{A}$, and we have seen what this entails in terms of reliance behavior for any value of $\mathcal{A}$ in \Cref{subsec:discerning}.

\section{Applying the Framework to Assess Interventions}\label{sec:interventions}

In the following, we leverage the presented framework to interpret the different effects that interventions exhibit on human reliance behavior and decision quality.
We start with analyses based on hypothetical interventions to demonstrate the relevance of disentangling effects on reliance behavior and accuracy in general.
Subsequently, we use our framework to assess and visualize the effects of interventions studied in prior work.

\subsection{Understanding the Effects of Interventions}\label{subsec:understanding-effects}
Our theoretical results and the visual framework can be used to assess empirical studies in AI-assisted decision-making and understand them better.
Any such empirical finding would be a static point in the colored rectangle, from which we can immediately infer interesting properties, such as the quantity and quality of reliance, the exact percentages of correct and wrong adherence and overrides, and the effectiveness of the human-in-the-loop in complementing the AI system.

\begin{figure}[ht]
    \centering
    \includegraphics[width=0.5\textwidth]{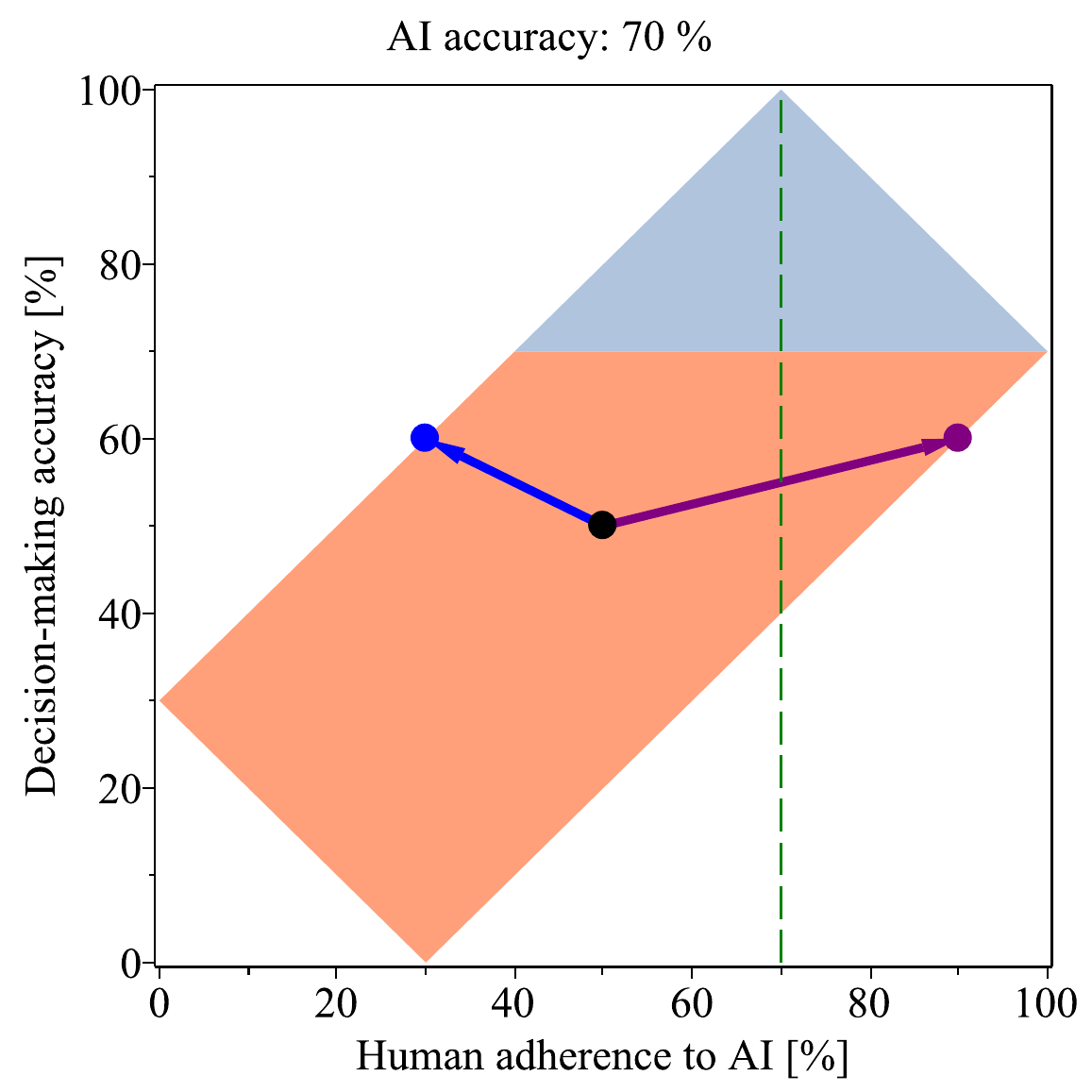}
    \caption{Visualizing the effects of two different interventions (blue \textcolor{blue}{$\bullet$} and purple \textcolor{Plum}{$\bullet$}) on reliance behavior and decision-making accuracy.}
    \label{fig:interventions}
\end{figure}

Another key usage of the framework is its ability to disentangle the effects of interventions, such as explanations or other means of decision support (see \shortciteA{lai2023towards} for an overview of such interventions).
For that, let us consider the following hypothetical example: through a randomized experiment, we have collected data where humans are relying on AI recommendations in the presence of two different types of explanations (blue \textcolor{blue}{$\bullet$} and purple \textcolor{Plum}{$\bullet$} dots) vs. a baseline without explanations (black dot $\bullet$).
We can think of these interventions as movements in our visual framework, as illustrated by the colored arrows in \Cref{fig:interventions}.
The black dot corresponds to a situation where a human-in-the-loop cannot discern correct and wrong AI recommendations and adheres to $\mathcal{A}=50\%$ AI recommendations.

In the case of the blue \textcolor{blue}{$\bullet$} intervention, we see that it leads to a decrease in the degree of adherence to AI recommendations, compared to the baseline ($\mathcal{A}=50\% \rightarrow \mathcal{A}=30\%$), but an increase in decision-making accuracy ($Acc_{final}=50\% \rightarrow Acc_{final}=60\%$) through a better reliance quality ($Q=0.5 \rightarrow Q=1$).
In the case of the purple \textcolor{Plum}{$\bullet$} intervention, we see the same effect with respect to accuracy but an entirely different effect on the reliance behavior: this intervention leads to an \textit{increase} in adherence to AI recommendations ($\mathcal{A}=50\% \rightarrow \mathcal{A}=90\%$).
At the same time, reliance quality drops from $Q=0.5$ to $Q=0$, which from \Cref{cor:over-reliance} we know corresponds to a situation of over-reliance where any of the 10\% overrides are of correct AI recommendations.
Overall, this implies that different interventions can have seemingly similar effects on decision quality but drastically different effects on reliance behavior.
Our framework enables us to disentangle these effects.

\subsection{Application to Empirical Studies}\label{sec:application}

To further emphasize its practical relevance, we now apply our framework to three empirical studies that investigate the effects of different interventions on reliance behavior and accuracy in AI-assisted decision-making.
The studies cover both binary and multi-class decision-making tasks.
We propose that the way we document and present the empirical findings can serve as a blueprint for researchers when reporting their own results.
 
\subsubsection{Introduction of Studies}

In \textbf{study~1} by \shortciteA{liu2021understanding}, the authors investigate how interactive explanations can be utilized to facilitate human-AI complementarity.
They also compare the performance of human and AI system for in-distribution and out-of-distribution setups originating from distribution shifts.
Overall, the authors observe significant performance differences on in-distribution and out-of-distribution data.
For interactive explanations, the authors observe mixed results; they suggest that interactive explanations can improve the perceived usefulness of AI assistance, but they may also reinforce human biases and lead to limited performance improvement. 

\textbf{Study~2} by \shortciteA{schoeffer2024explanations} assesses the effects of feature-based explanations on the fairness of AI-assisted decisions, through the mediating role of reliance behavior.
The research shows that feature-based explanations may foster or hinder fairness, depending on which type of features they highlight.
Moreover, this study suggests that humans override $(i)$ more AI recommendations when explanations highlight gendered features, and $(ii)$ fewer AI recommendations when only task-relevant features are highlighted.
At the same time, the authors find that feature-based explanations do not enhance people's ability to execute correct vs. wrong overrides compared to a baseline scenario without explanations.

In \textbf{study~3} by \shortciteA{tejeda2022ai}, the authors compare differences in reliance behavior based on whether or not a human-in-the-loop provides independent decisions before they are provided with AI assistance (i.e., \textit{sequential} vs. \textit{concurrent} decision-making).
This work develops a cognitive model to infer the latent reliance strategy of humans on AI assistance without asking them to make an independent decision beforehand.
The predictions of this cognitive model are then validated through behavioral experiments for a multi-class classification task.
Specifically, participants had to classify everyday images, such as different animals or types of vehicles.

\subsubsection{Implementation}

We apply our proposed framework to these studies to visualize the effects of the respective interventions and to gain a deeper understanding of how the interventions affect the ability of humans to discern between correct and wrong AI recommendations.
To this end, we implemented a user interface that allows to customize the visual framework to any empirical study requiring only information on the AI accuracy ($Acc_{AI}$), the level of adherence ($\mathcal{A}$), and the final decision-making accuracy ($Acc_{final}$).\footnote{The tool is available at \url{https://github.com/jhnnsjkbk/accuracy-reliance}.}
We leverage the developed tool to generate the figures in this section.
Moreover, in addition to the visualization, our implementation solves the following system of equations to determine the proportions of correct and wrong overrides as well as correct and wrong adherence:
\begin{align*}
 \mathcal{A}_{correct} + \mathcal{O}_{correct} &= Acc_{final} \\ 
 \mathcal{A}_{correct} + \mathcal{A}_{wrong}  &= \mathcal{A} \\ 
 \mathcal{A}_{correct} + \mathcal{O}_{wrong}  &= Acc_{AI} \\ 
 \mathcal{A}_{correct} + \mathcal{A}_{wrong} + \mathcal{O}_{correct} + \mathcal{O}_{wrong}  &= 100\%.
\end{align*}
Note that, at the moment, the tool is primarily developed for binary decision-making tasks. 
Finally, our implementation calculates the quality of reliance, $Q$.
As defined in \Cref{eq:quality} in \Cref{sec:quality-of-reliance}, the quality of reliance represents the ability of the human-in-the-loop to discern correct from wrong AI recommendations given a level of adherence $\mathcal{A}$.
Importantly, this measure is only dependent on $\mathcal{A}$ and the observed decision-making accuracy ($Acc_{final}$), but it is not influenced by the AI accuracy---facilitating comparisons of empirical findings across different employed AI models and thus across empirical studies.

We briefly explain the data aggregation process for the three studies.
For \textbf{study~1}, we focus on the \textit{in-distribution} condition on the ICPSR dataset and obtain scores for the two interventions called \textit{static/static} and \textit{interactive/interactive}.
We obtain the adherence levels per intervention directly from the respective paper by \shortciteA{liu2021understanding}: $\mathcal{A}=85.9\%$ (\textit{static/static}) and $\mathcal{A}=87.9\%$ (\textit{interactive/interactive}).
Similarly, we obtain the values for decision-making accuracy as $Acc_{final}=61.5\%$ (\textit{static/static}) and $Acc_{final}=60.7\%$ (\textit{interactive/interactive}).
From that, and the reported \textit{negative} accuracy gains from interventions, we infer the AI accuracy as follows: $Acc_{AI} = 61.5\% - (-6.2)\% = 67.7\%$ (\textit{static/static}) and $Acc_{AI} = 60.7\% - (-6.3)\% = 67.0\%$ (\textit{interactive/interactive}).
Thus, we obtain an average AI accuracy of $Acc_{AI} = 67.4\%$.
As we have access to the raw data for \textbf{study~2}, we directly use this data to determine accuracy and adherence values.
Similarly, for \textbf{study~3}, we are able to directly infer the aggregated values for AI accuracy, adherence, and the final decision-making accuracy across the three conditions \textit{weak AI}, \textit{medium AI}, and \textit{strong AI} from the publicly available raw data.

\subsubsection{Analysis of Studies}

\begin{figure*}[t]
    \centering
    \begin{subfigure}[t]{0.48\textwidth}
        \centering
        \includegraphics[width=1\textwidth]{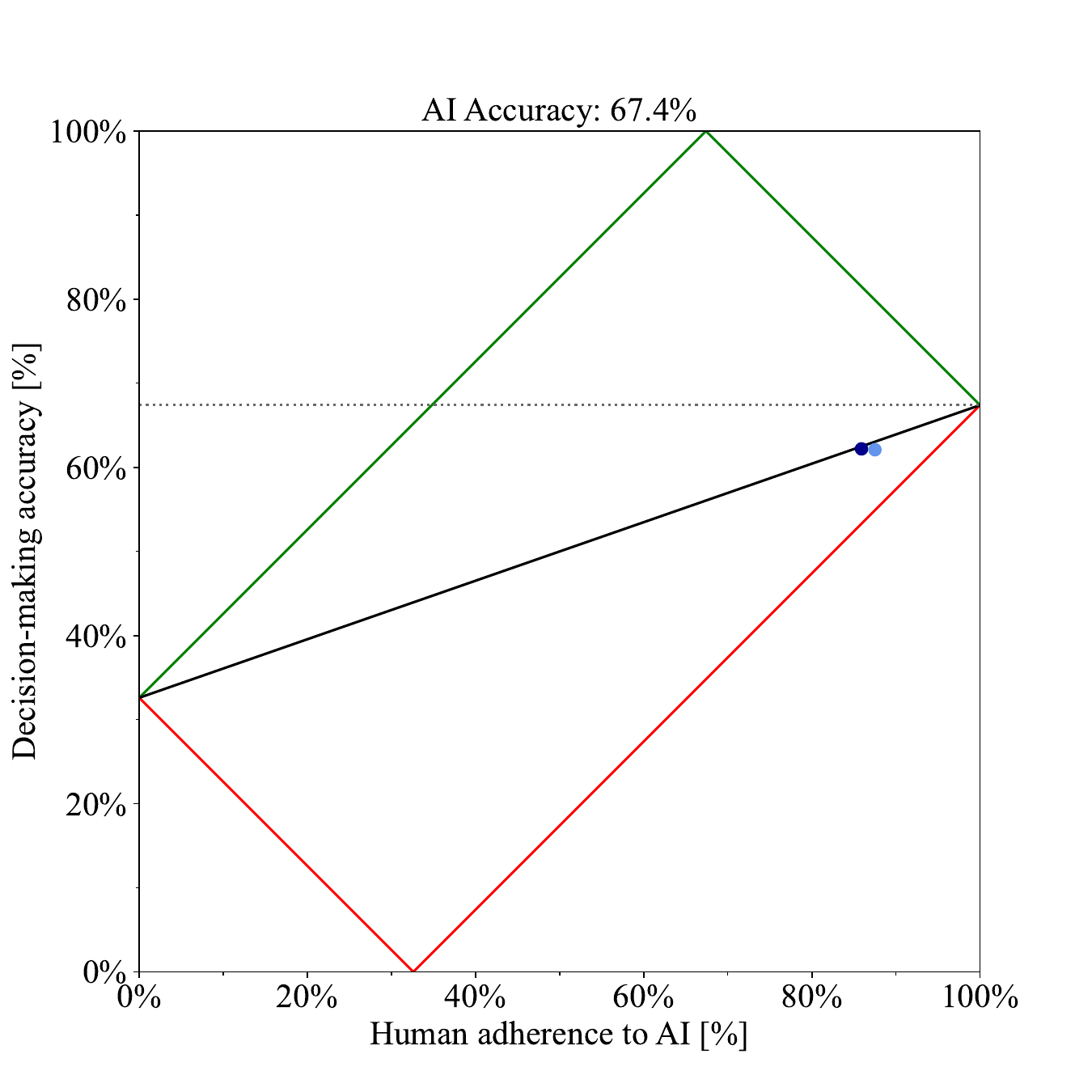}
        \caption{\textbf{Study~1}~\shortcite{liu2021understanding}.}
    \end{subfigure}
    \hfill
    \begin{subfigure}[t]{0.48\textwidth}
        \centering
        \includegraphics[width=1\textwidth]{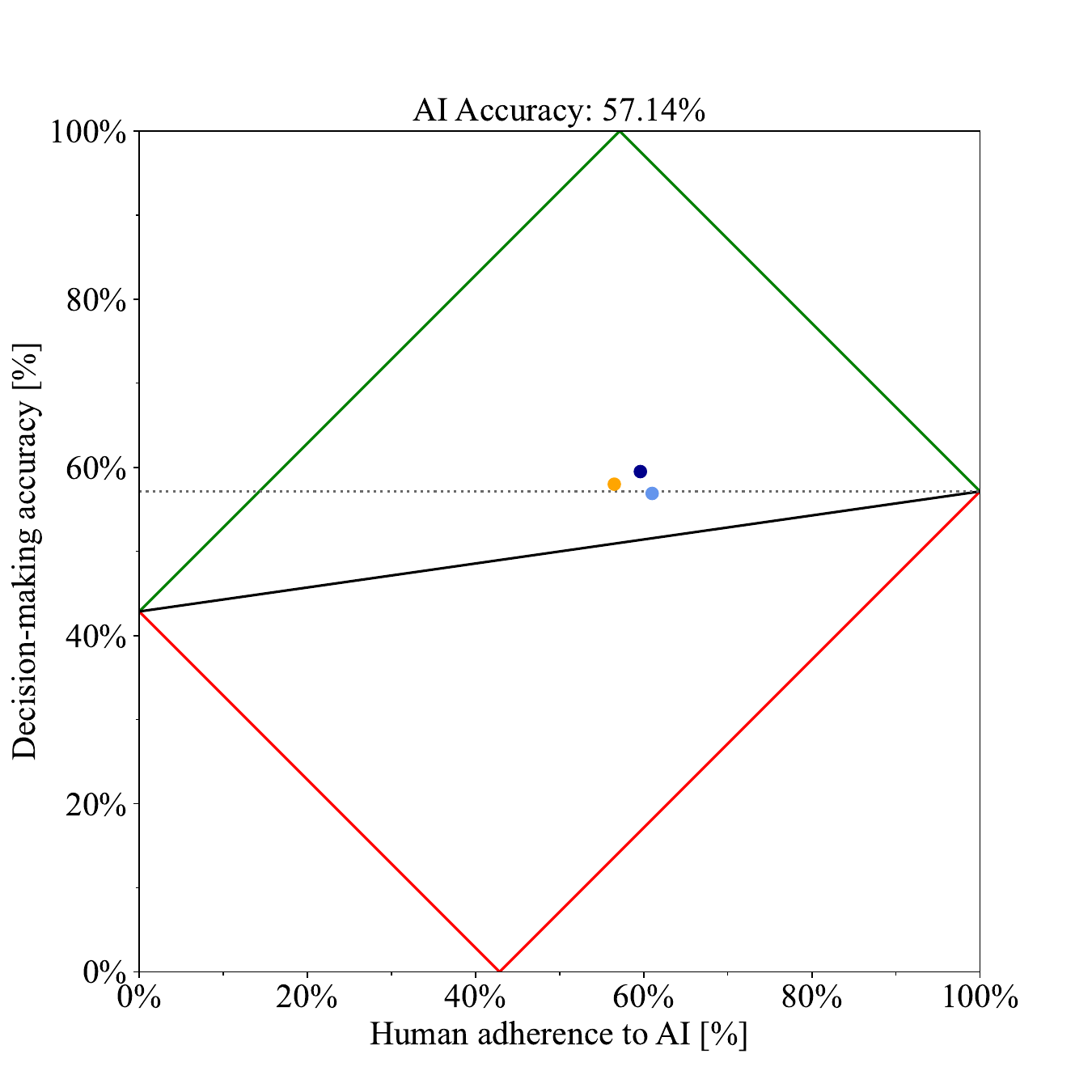}
        \caption{\textbf{Study~2}~\shortcite{schoeffer2024explanations}.}
    \end{subfigure}
    \hfill
    \begin{subfigure}[t]{0.48\textwidth}
        \centering
        \includegraphics[width=1\textwidth]{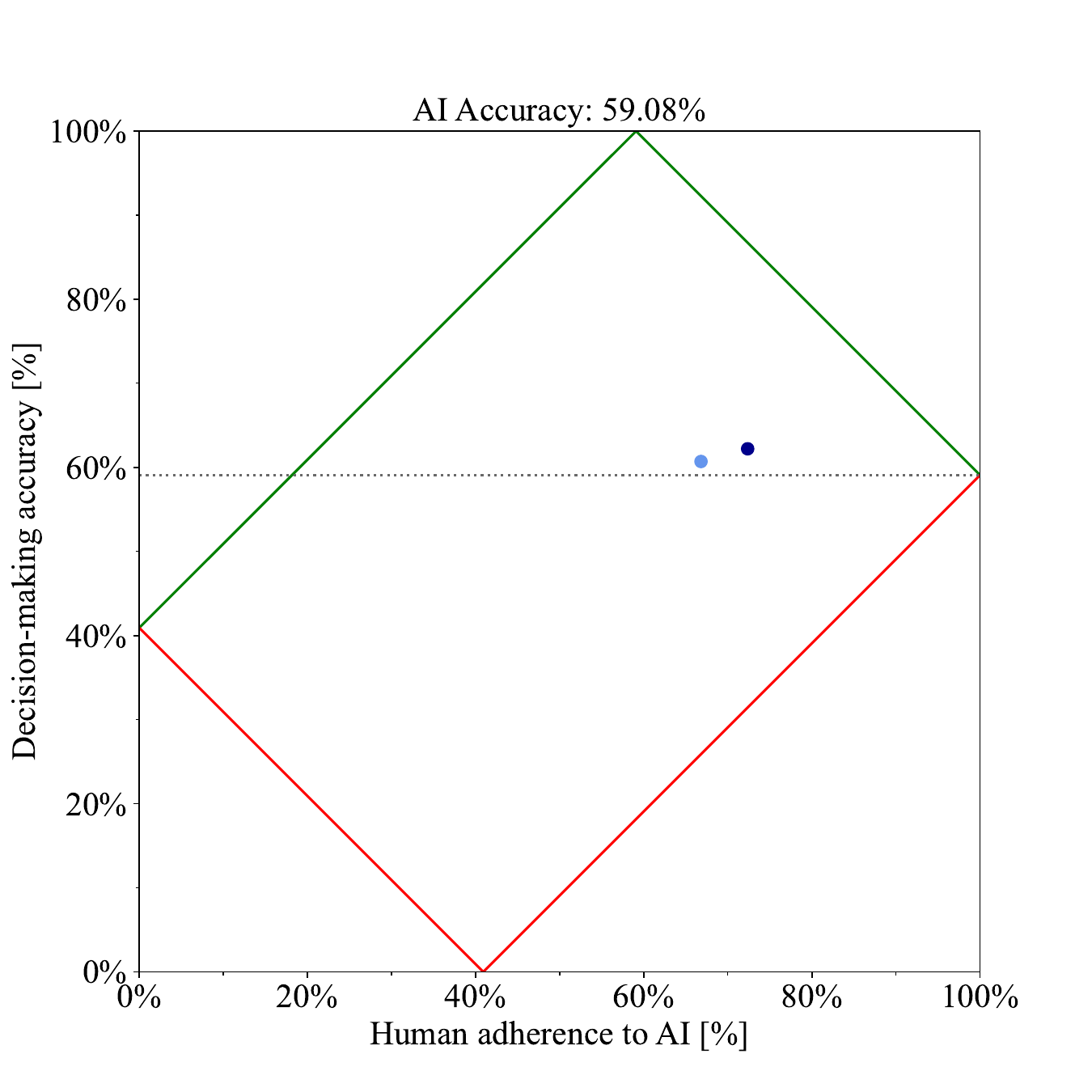}
        \caption{\textbf{Study~3}~\shortcite{tejeda2022ai}.}
    \end{subfigure}
    \caption{Visualization of empirical findings on the effects of different interventions (colored dots) on human reliance and decision-making accuracy using our proposed framework. Note that we do not show the level of random guessing (black line) in the multi-class setting (c).}\label{fig:application}
\end{figure*}

In the following, we visualize the empirical findings from the selected studies with the help of our framework and present the results in \Cref{fig:application}.
The framework summarizes the key findings within a single visualization and extends them by additional observations.
For instance, participants in \textbf{study~1} were performing worse than the AI alone in both cases (with static and interactive explanations).
Notably, the participants' reliance behavior was similar to that of a random guess.
The different types of explanations resulted both in very similar reliance behavior and decision-making accuracy.
The degree of adherence to AI recommendations was slightly higher for intervention~2 (interactive explanations; light blue~\textcolor{Cerulean}{$\bullet$}) compared to intervention~1 (static explanations; dark blue \textcolor{Blue}{$\bullet$}).
However, in both cases, the level of adherence was significantly higher than the optimal level of adherence of 67.4\%, as can be seen in \Cref{fig:application} (a). 
This means that participants \textit{over-relied} on AI recommendations.

This is different in \textbf{study~2}, visualized in \Cref{fig:application} (b), where both the baseline condition and the interventions resulted in levels of adherence that are more aligned with the optimal level of adherence of 57.1\%.
Compared to the baseline condition (dark blue~\textcolor{Blue}{$\bullet$}), intervention~2 (\textit{gendered}; orange \textcolor{YellowOrange}{$\bullet$}) reduced the reliance quantity while intervention~1 (\textit{task-relevant}; light blue~\textcolor{Cerulean}{$\bullet$}) increased it.
We observe that intervention~2 resulted in a level of adherence closer to the optimum.
That is, adherence was reduced from 59.6\% in the baseline condition to 56.5\% based on intervention~2, while the optimal level of adherence is given by 57.1\%.
Although both intervention~1 and 2 resulted in a reduced decision-making accuracy compared to the baseline, our framework shows that intervention~2 improves over intervention~1 both in \textit{quantity} of reliance and in \textit{quality}.

For \textbf{study~3}, we observe that sequential decision-making (light blue~\textcolor{Cerulean}{$\bullet$}) resulted in a reduced decision-making accuracy compared to concurrent decision-making (dark blue~\textcolor{Blue}{$\bullet$}), as shown in \Cref{fig:application} (c).
This is similar to the effect of intervention~2 in \textbf{study~2}.
Additionally, adherence to AI recommendations decreased when moving from concurrent to sequential decision-making.
Interestingly, this moves the degree of adherence closer to  the optimal level of 59.1\%.
Hence, sequential decision-making improved the level of adherence but reduced the quality of reliance.

After having obtained a general overview of the empirical findings through \Cref{fig:application}, we further utilize our framework to gain a more detailed understanding of the reliance behavior induced by different interventions.
We summarize these insights per intervention for \textbf{study~1} and \textbf{study~2} in \Cref{tab:application-v1}.\footnote{For multi-class decisions, our framework does not currently allow inferring whether adherence and overriding was correct or wrong.}
First, based on \Cref{eq:def_relationships} in \Cref{sec:the_interdependence}, we derive information on the proportions of correct and wrong adherence as well as correct and wrong overrides.
Overall, we observe that interventions affect both the quantity and quality of reliance across studies. 
For instance, as noted, the level of adherence is significantly higher in \textbf{study~1} than in \textbf{study~2}; however, the ratio of correct adherence to overall adherence is comparable: for intervention~1 of \textbf{study~1}, we have $\mathcal{A}_{correct}/\mathcal{A}=66.8\%$, compared to $63.9\%$ in the baseline condition of \textbf{study~2}.
This implies that when participants followed AI recommendations, they relied on incorrect advice in roughly one-third of the cases.
We may interpret the ratio $\mathcal{A}_{correct}/\mathcal{A}$ as the \textit{precision} of adherence, and we further explore this and related metrics in \Cref{sec:f-scores}. 

We also see differences in overriding behavior across studies. 
Interestingly, participants in \textbf{study~1} overrode more correct than wrong AI recommendations (i.e., $\mathcal{O}_{wrong} > \mathcal{O}_{correct}$), suggesting a deficient ability to distinguish correct from wrong AI advice.
\begin{table}[t]
    \centering
    \begin{tabular}{l l | l l}
    \toprule
    \multicolumn{2}{l|}{\textbf{Study~1} \shortcite{liu2021understanding}} & \multicolumn{2}{l}{\textbf{Study~2} \shortcite{schoeffer2024explanations}} \\
    \midrule
        \textbf{Intervention 1} & $\mathcal{A}_{correct}= 57.4\%$ & \textbf{Baseline} & $\mathcal{A}_{correct}= 38.1\%$ \\
        (dark blue \textcolor{Blue}{$\bullet$}) & $\mathcal{A}_{wrong}= 28.5\%$ & (dark blue \textcolor{Blue}{$\bullet$}) & $\mathcal{A}_{wrong}= 21.5\%$ \\
        & $\mathcal{O}_{correct}= 4.1\%$ & & $\mathcal{O}_{correct}= 21.4\%$ \\
        & $\mathcal{O}_{wrong}= 10.0\%$ & & $\mathcal{O}_{wrong}= 19.0\%$ \\
        & $Q=0.29$ & & $Q=0.53$ \\
        \midrule
        \textbf{Intervention 2} & $\mathcal{A}_{correct}= 58.0\%$ & \textbf{Intervention 1} & $\mathcal{A}_{correct}= 37.5\%$ \\
        (light blue \textcolor{Cerulean}{$\bullet$}) & $\mathcal{A}_{wrong}= 29.9\%$ & (light blue \textcolor{Cerulean}{$\bullet$}) & $\mathcal{A}_{wrong}= 23.5\%$ \\
        & $\mathcal{O}_{correct}= 2.7\%$ & & $\mathcal{O}_{correct}= 19.4\%$ \\
        & $\mathcal{O}_{wrong}= 9.4\%$ & & $\mathcal{O}_{wrong}= 19.6\%$ \\
        & $Q=0.22$ & & $Q=0.50$ \\
        \midrule 
         & 
         & \textbf{Intervention 2} & $\mathcal{A}_{correct}= 35.8\%$ \\
         & & (orange \textcolor{YellowOrange}{$\bullet$}) & $\mathcal{A}_{wrong}= 20.7\%$ \\
        & & & $\mathcal{O}_{correct}= 22.2\%$ \\
        & & & $\mathcal{O}_{wrong}= 21.3\%$ \\
        & & & $Q=0.52$\\
    \bottomrule
    \end{tabular}
    \caption{Empirical findings from \shortciteA{liu2021understanding} and \shortciteA{schoeffer2024explanations} interpreted through our framework.}
    \label{tab:application-v1}
\end{table}
This is also reflected in the quality of reliance, $Q$, which we calculate based on \Cref{eq:quality} in \Cref{sec:quality-of-reliance}.\footnote{Recall that our measure of reliance quality, $Q$, is independent of AI accuracy, enabling meaningful comparisons of empirical findings across different studies.}
Here, we observe that the quality of reliance is significantly lower in \textbf{study~1} compared to \textbf{study~2}.
Since these differences in reliance quality persist across interventions, this suggests that general factors such as task complexity or the participants' backgrounds may vary between the two studies.

\section{Extensions of the Framework}\label{sec:extensions}
In the preceding sections, we outlined the key aspects of our proposed framework and applied it to study the effects of interventions.
In this section, we derive relevant extensions of the framework.
Specifically, we leverage our framework to estimate the likelihood of achieving human-AI complementarity (\Cref{sec:estimating_likelihood}), and we introduce a possible evaluation of decision quality beyond accuracy based on $F_{\beta}$-scores (\Cref{sec:f-scores}).

\subsection{Estimating the Likelihood of Achieving Complementarity}\label{sec:estimating_likelihood}
It is sometimes important to estimate the likelihood of realizing a decision-making accuracy that is greater than the AI accuracy, i.e., $Acc_{final} > Acc_{AI}$, meaning an ability of the human-in-the-loop to complement the AI.
For that, let us recall from \Cref{cor:under-rely} that such complementarity cannot be achieved if the human under-relies on AI recommendations at $\mathcal{A} \leq 2\cdot Acc_{AI}-100\%$.
For instance, when we have an AI accuracy of $Acc_{AI}=70\%$, this adherence threshold is at $\mathcal{A}=40\%$ (see, e.g., \Cref{fig:framework_dev} (a) in \Cref{subsec:visual_framework}).
That is, the human-in-the-loop can theoretically improve the decision-making accuracy over the initial AI accuracy if $(i)$ they adhere to more than 40\% of AI recommendations (or, equivalently, override \textit{less than} 60\%), and $(ii)$ their reliance quality is sufficiently high.

\begin{figure}
    \centering
    \includegraphics[width=\textwidth]{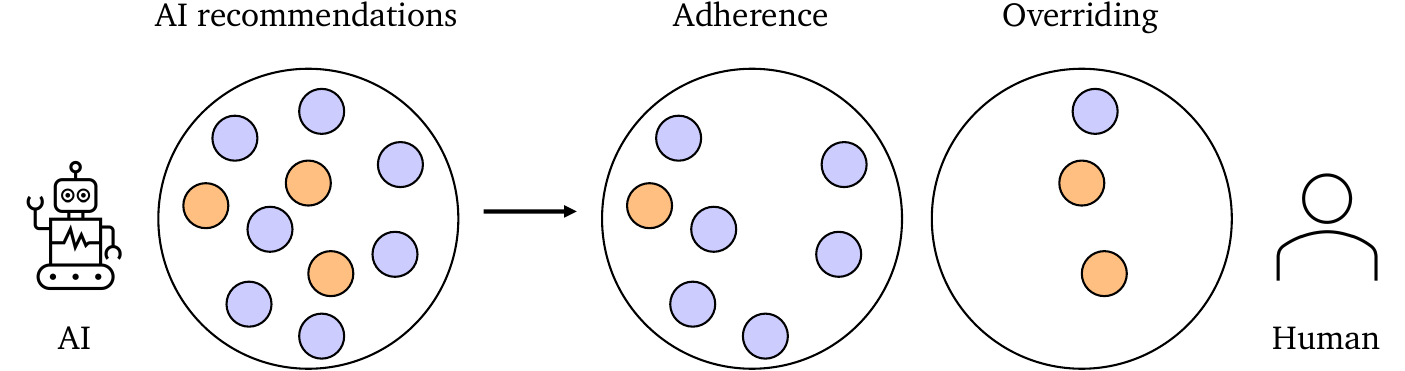}
    \caption{An exemplary human reliance on AI recommendations illustrated as marbles in an urn. In this example, we have $n=10$, $Acc_{AI}=70\%$, $\mathcal{A}=70\%$, and a resulting decision-making accuracy of $Acc_{final}=80\%$, consisting of 6 instances of correct adherence and 2 instances of correct overriding. Blue marbles are correct AI recommendations; orange marbles are wrong.}
    \label{fig:urn_example}
\end{figure}

To estimate the likelihood of achieving complementarity (i.e., $Acc_{final} > Acc_{AI}$), we can simulate the human reliance behavior as an urn model, where we have $n\in \mathbf{N}$ marbles that correspond to individual AI recommendations.
The colors of marbles indicate their correctness (blue vs. orange), and drawing a marble corresponds to adhering to a given AI recommendation.
This is illustrated for the case of $n=10$ in \Cref{fig:urn_example}.
Here, an AI issues $n=10$ recommendations, 3 of which are wrong ($Acc_{AI}=70\%$); i.e., we have 3 orange and 7 blue marbles in the urn.
In this example, the human-in-the-loop then selects 7 marbles to indicate adherence to the respective AI recommendations ($\mathcal{A}=70\%$), one of which is wrong. The remaining 3 marbles are left in the urn and imply that the corresponding AI recommendations (2 wrong and 1 correct) are overridden.

If the human-in-the-loop cannot distinguish correct from wrong AI recommendations, which means their reliance behavior is \textit{independent} of the correctness of an AI recommendation, we expect a decision-making accuracy of
\begin{equation*}
    Acc_{final}(\mathcal{A})=(100\%-Acc_{AI}) + (2\cdot Acc_{AI}-100\%) \cdot \mathcal{A},
\end{equation*}
as stated in \Cref{prop:discern}.
We also know from \Cref{cor:lower} that this value is always lower than the AI accuracy, except for the case of $\mathcal{A}=100\%$ where they are equal.
However, even if the human-in-the-loop cannot distinguish correct from wrong AI recommendations, a decision-making accuracy of $Acc_{final} > Acc_{AI}$ may in practice be realized by sheer luck---especially when $n$ is small.
This has important implications for the interpretation of empirical findings.
We now simulate different scenarios and infer the likelihood of observing complementarity.

For that, we model the human reliance behavior as sampling without replacement and without order from an urn that contains a finite number $n$ of blue ($b\in \mathbf{N}$) and orange ($o\in \mathbf{N}$) marbles,\footnote{Note that the values of $b$ and $o$ follow immediately from a given AI accuracy: $b=\lfloor Acc_{AI}\cdot n \rceil$; $o=n-b$, where we use $\lfloor \cdot \rceil$ to denote rounding to the nearest integer.} with $n=b+o$.
The marbles are indistinguishable to the human.
We know that the probability of drawing $k\in \mathbf{N}$ blue marbles out of $a=\lfloor\mathcal{A}\cdot n\rceil \in \mathbf{N}$ attempts is described by a hypergeometric distribution with parameters $(n,b,a)$ and probability mass function
\begin{equation}\label{eq:binom}
    \Pr(k)= \frac{\binom{b}{k}\binom{n-b}{a-k}}{\binom{n}{a}}.
\end{equation}
Coming back to our example from \Cref{fig:urn_example}, the relevant values are: $n=10$, $b=7$, $o=3$, and $a=7$.
Recall that $a$ is fixed ex ante.
In order to achieve complementarity, the human-in-the-loop would have to choose either $k=6$ or $k=7$ blue marbles.
Hence, the probability of observing complementarity given the marbles are not distinguishable is
\begin{equation*}
    \Pr(k\geq6)=\Pr(k=6)+\Pr(k=7)=\frac{\binom{7}{6}\binom{3}{1}}{\binom{10}{7}} + \frac{\binom{7}{7}\binom{3}{0}}{\binom{10}{7}} = \frac{7}{40} + \frac{1}{120} = \frac{11}{60} \approx 18.3\%. 
\end{equation*}
Hence, the likelihood of observing $Acc_{final} > Acc_{AI}$ by sheer chance in the given example is approximately 18.3\%.
This is also reflected in \Cref{tab:ablation}, which contains the likelihood of randomly achieving complementarity for all relevant cases of $n=10$.

Based on this example, we now conduct a follow-up study to understand the impacts of $Acc_{AI}$ and $\mathcal{A}$ on the likelihood of randomly observing complementarity.
For that, we first establish in general how many blue marbles (i.e., correct adherence) the human-in-the-loop needs to choose in order to achieve complementarity.

\begin{mdframed}[nobreak=true]
\begin{proposition}\label{prop:marbles}
    Given $\mathcal{A} \geq 2\cdot Acc_{AI}-100\%$, achieving (strict) complementarity, i.e., $Acc_{final} > Acc_{AI}$, is equivalent to $(i)$ and $(ii)$:
    \begin{itemize}
        \item[$(i)$] $0 \leq \mathcal{O}_{wrong} < \mathcal{O}_{correct}$
        \item[$(ii)$] $Acc_{AI} - 50\% + 0.5\cdot \mathcal{A} < \mathcal{A}_{correct} \leq \min\{Acc_{AI},\mathcal{A}\}$.
    \end{itemize}
\end{proposition}
\end{mdframed}

\Cref{prop:marbles} follows from \Cref{cor:under-rely} as well as \Cref{eq:def_relationships} in \Cref{sec:the_interdependence}.
Part~$(i)$ means that the human-in-the-loop \textit{must} override at least one AI recommendation, and more overrides must be of wrong recommendations vs. correct ones.
Part~$(ii)$ implies that given an AI accuracy $Acc_{AI}$ and a sufficiently high level of adherence $\mathcal{A}$, the lower bound on the share of correct adherence that still leads to complementarity is $Acc_{AI} - 50\% + 0.5\cdot \mathcal{A}$.
For instance, if we have an AI accuracy of $Acc_{AI}=70\%$ and a level of adherence $\mathcal{A}=70\%$, then more than 55\% of this adherence must be to correct AI recommendations (i.e., $\mathcal{A}_{correct} > 55\%$); which directly implies $\mathcal{A}_{wrong} < 15\%$, $\mathcal{O}_{wrong} < 15\%$, and $\mathcal{O}_{correct} > 15\%$.
\Cref{prop:marbles} also allows us to determine the values of $k$ in \Cref{eq:binom} for which complementarity is achieved.

\begin{table}[t]
    \centering
    \begin{tabular}{c|cccccccc}
    \toprule
        $\mathcal{A}$ & 30\% & 40\% & 50\% & 60\% & 70\% & 80\% & 90\% \\
        $Acc_{AI}$ & ($a=3$) & ($a=4$) & ($a=5$) & ($a=6$) & ($a=7$) & ($a=8$) & ($a=9$) \\
        \midrule
        60\% ($b=6$) & 16.7\% & 7.1\% & 26.2\% & 11.9\% & 33.3\% & 13.3\% & 40.0\% \\
        70\% ($b=7$) & & & 8.3\% & 3.3\% & 18.3\% & 2.2\% & 30.0\% \\
        80\% ($b=8$) & & & & & 6.7\% & 2.2\% & 4.4\% \\ 
        90\% ($b=9$) & & & & & & & 10.0\% \\
    \bottomrule
    \end{tabular}
    \caption{The likelihood of observing complementarity (i.e., $Acc_{final} > Acc_{AI}$) by chance, as a function of $Acc_{AI}$ and $\mathcal{A}$, given $n=10$. Empty cells correspond to cases where no complementarity is possible (see \Cref{cor:under-rely}).}
    \label{tab:ablation}
\end{table}

\begin{figure}[ht]
    \centering
    \begin{minipage}[t]{0.48\textwidth}
        \centering
        \includegraphics[width=0.95\textwidth]{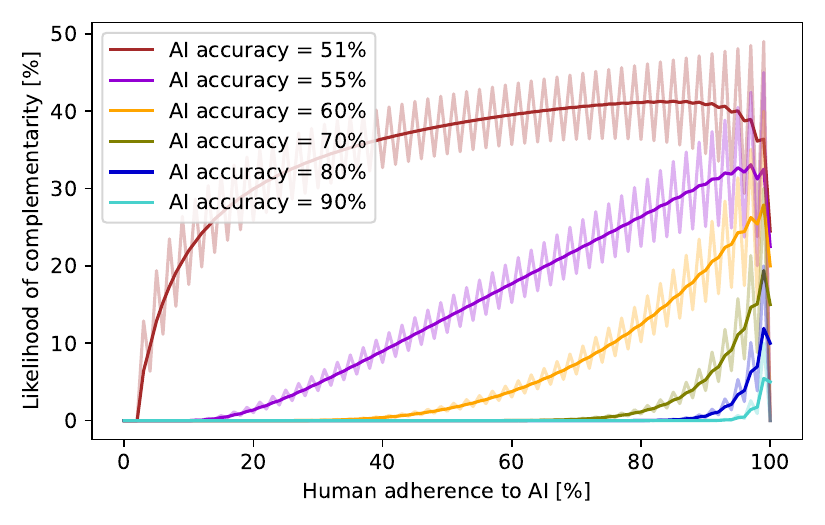}
        \caption{Likelihood of achieving complementarity (i.e., $Acc_{final} > Acc_{AI}$) when the human-in-the-loop cannot distinguish correct from wrong AI recommendations, as a function of the degree of human adherence, for different levels of AI accuracy. We also include moving averages.}
        \label{fig:compl_over_adh}
    \end{minipage}\hfill
    \begin{minipage}[t]{0.48\textwidth}
        \centering
        \includegraphics[width=0.95\textwidth]{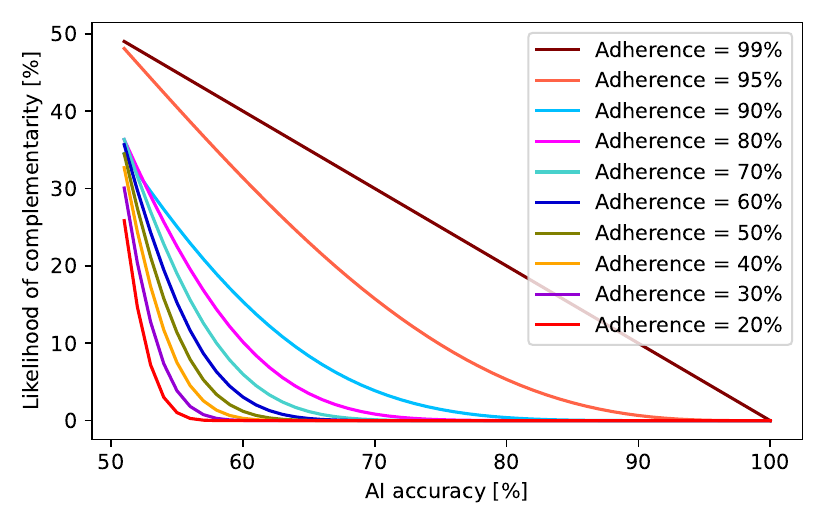}
        \caption{Likelihood of achieving complementarity when the human-in-the-loop cannot distinguish correct from wrong AI recommendations, as a function of the AI accuracy, for different levels of human adherence to AI recommendations.}
        \label{fig:compl_over_AI}
    \end{minipage}\hfill
\end{figure}

We can now analyze the effects of $Acc_{AI}$ and $\mathcal{A}$ on the likelihood of observing complementarity by chance.
For clarity of exposition as well as computational tractability, we focus on $n=100$.
This means that our hypothetical urn contains 100 marbles, and we can treat one marble as one percentage point.
The results are visualized in \Cref{fig:compl_over_adh,fig:compl_over_AI}.
First, in \Cref{fig:compl_over_adh}, we see that the likelihood of achieving complementarity increases in the degree of human adherence to AI recommendations, and that the likelihood is generally lower as the AI accuracy increases. Because the likelihoods are oscillating, we also provide moving averages to enhance readability.
Importantly, in cases where the human-in-the-loop adheres to AI recommendations very often, the likelihood of achieving complementarity by chance becomes significant.
For instance, when we have $Acc_{AI}=60\%$ and an adherence of $\mathcal{A}=95\%$, this likelihood is approximately 31.4\%.
In other words, if humans cannot distinguish correct from wrong AI recommendations, they may increase their chances of achieving $Acc_{final} > Acc_{AI}$ by just adhering to more AI recommendations;\footnote{Recall that we assume $Acc_{AI}>50\%$, i.e., the AI performs better than chance.} and these chances can become relatively high---especially when $Acc_{AI}\ll100\%$.
In such cases, it may sometimes seem that a human-in-the-loop is adding value to the system when in reality they are just over-relying on AI advice with no increased ability to distinguish correct from wrong AI recommendations.
In \Cref{fig:compl_over_AI}, we further see that the likelihood of randomly achieving complementarity decreases as the AI accuracy increases, and this decrease happens faster for low levels of adherence $\mathcal{A}$.

\subsection{Assessing Decision Quality with $F_{\beta}$-Scores}\label{sec:f-scores}

In the previous sections, we have primarily assessed decision quality using the accuracy metric, as it is the most widely adopted metric and allows for a clear and concise overview of our framework.
However, the interdependence of reliance behavior and decision quality does not require an assessment based on accuracy but can be extended to other metrics that are a function of correct and wrong decisions, which we detail in the following.

When the underlying class distribution of the data is imbalanced, the accuracy metric is prone to misleading interpretations~\shortcite{maratea2014adjusted}.
For this reason, more sophisticated metrics have evolved for evaluating the quality of predictions.
One such metric, which involves a weighted combination of the \textit{precision} and \textit{recall} metrics~\shortcite{goutte2005probabilistic}, is formally defined for $\beta \geq 0$ as follows:
\begin{equation*}
    F_{\beta} = (1+\beta^2) \cdot \frac{precision \cdot recall}{(\beta^2 \cdot precision) + recall},
\end{equation*}
where
\begin{equation*}
    precision = \frac{\textit{true positive}}{\textit{true positive} + \textit{false positive}},\quad recall = \frac{\textit{true positive}}{\textit{true positive} + \textit{false negative}}.
\end{equation*}
The special case of $\beta=1$, called $F_{1}$-score, defines a harmonic mean of precision and recall and is frequently used to assess the performance of machine learning models and associated systems on imbalanced datasets.
The $F_1$-score is defined as
\begin{equation*}
    F_{1} = \frac{2 \cdot precision \cdot recall}{precision + recall}.
\end{equation*}

In the following, we demonstrate how the $F_{1}$-score can serve as a quality metric in AI-assisted decision-making and how it relates to reliance behavior.
While, in general, we could compute the $F_{1}$-score also for the AI performance (alongside the final decision quality), it is not necessary for the subsequent considerations.
Instead, we reinterpret the adherence to correct AI recommendations ($\mathcal{A}_{correct}$) as
\textit{true positive}, the overriding of wrong AI recommendation ($\mathcal{O}_{correct}$) as \textit{true negative}, the adherence to
wrong AI recommendations ($\mathcal{A}_{wrong}$) as \textit{false positive}, and the overriding of correct AI recommendations ($\mathcal{O}_{wrong}$)
as \textit{false negative}.
With these definitions in place, we can interpret \textit{precision} as the fraction of correct AI recommendations among those that a human adheres to ($\mathcal{A}_{correct}/\mathcal{A}$); and we can think of \textit{recall} as the fraction of correct AI recommendations that a human adheres to ($\mathcal{A}_{correct}/Acc_{AI}$).

For an illustration of the interdependence of reliance and decision quality based on the $F_1$-score, let us reconsider the example from \Cref{fig:sys_acc_examples} in \Cref{sec:motivational}, with an AI accuracy of $Acc_{AI}=70\%$.
At an adherence level of $\mathcal{A}=100\%$, the human-in-the-loop is correctly adhering to the AI recommendations in 7 out of 10 cases (i.e., \textit{true positives}) and wrongly adhering to the recommendations in 3 out of 10 cases (i.e., \textit{false positives}).
Thus, for an AI accuracy of $Acc_{AI}=70\%$ and adherence $\mathcal{A}=100\%$, the corresponding final decision quality in terms of the $F_{1}$-score is

\begin{equation*}
    F_1 = \frac{2 \cdot 0.7 \cdot 1}{0.7 + 1} = 82.35\%.
\end{equation*}

\begin{figure}[ht]
    \centering
    \begin{minipage}[t]{0.48\textwidth}
        \centering
        \includegraphics[width=0.95\textwidth]{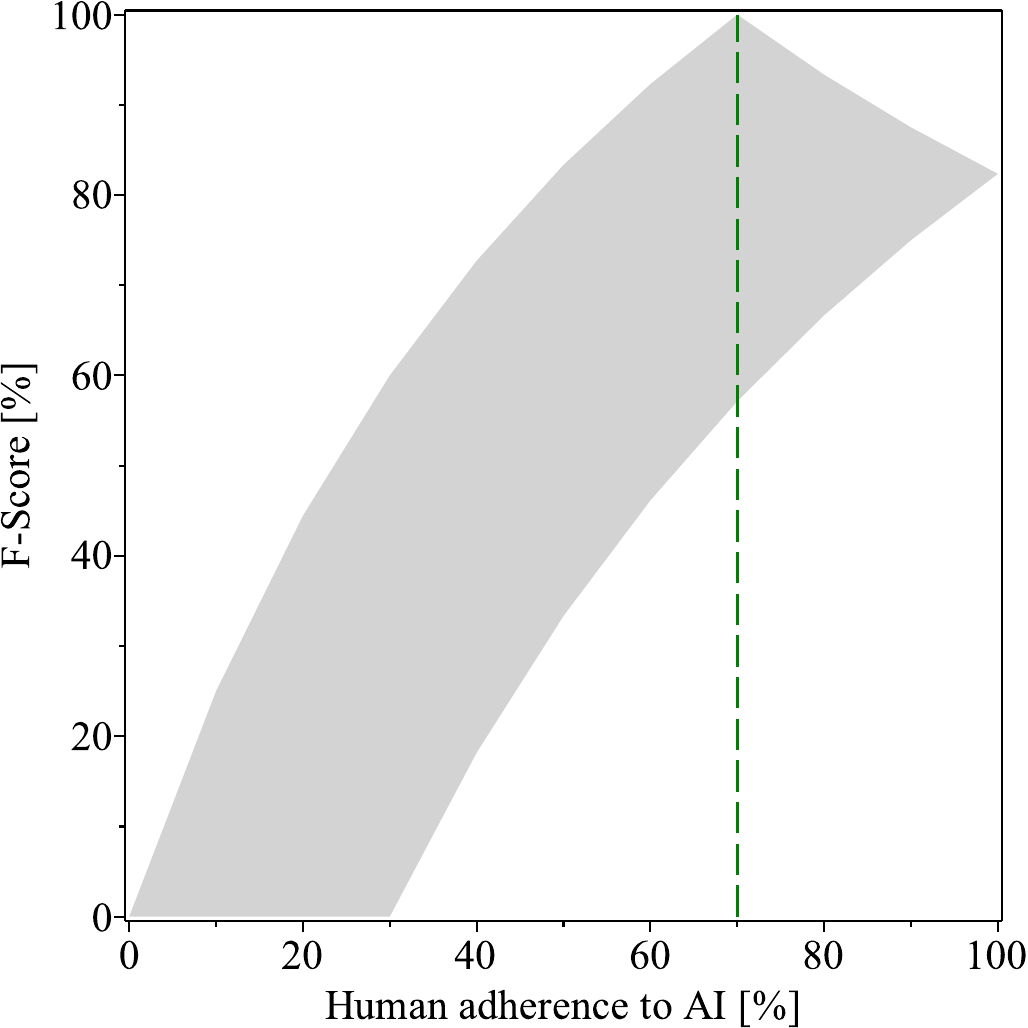}
        (a) $F_{1}$-Score
    \end{minipage}\hfill
    \begin{minipage}[t]{0.48\textwidth}
        \centering
        \includegraphics[width=0.95\textwidth]{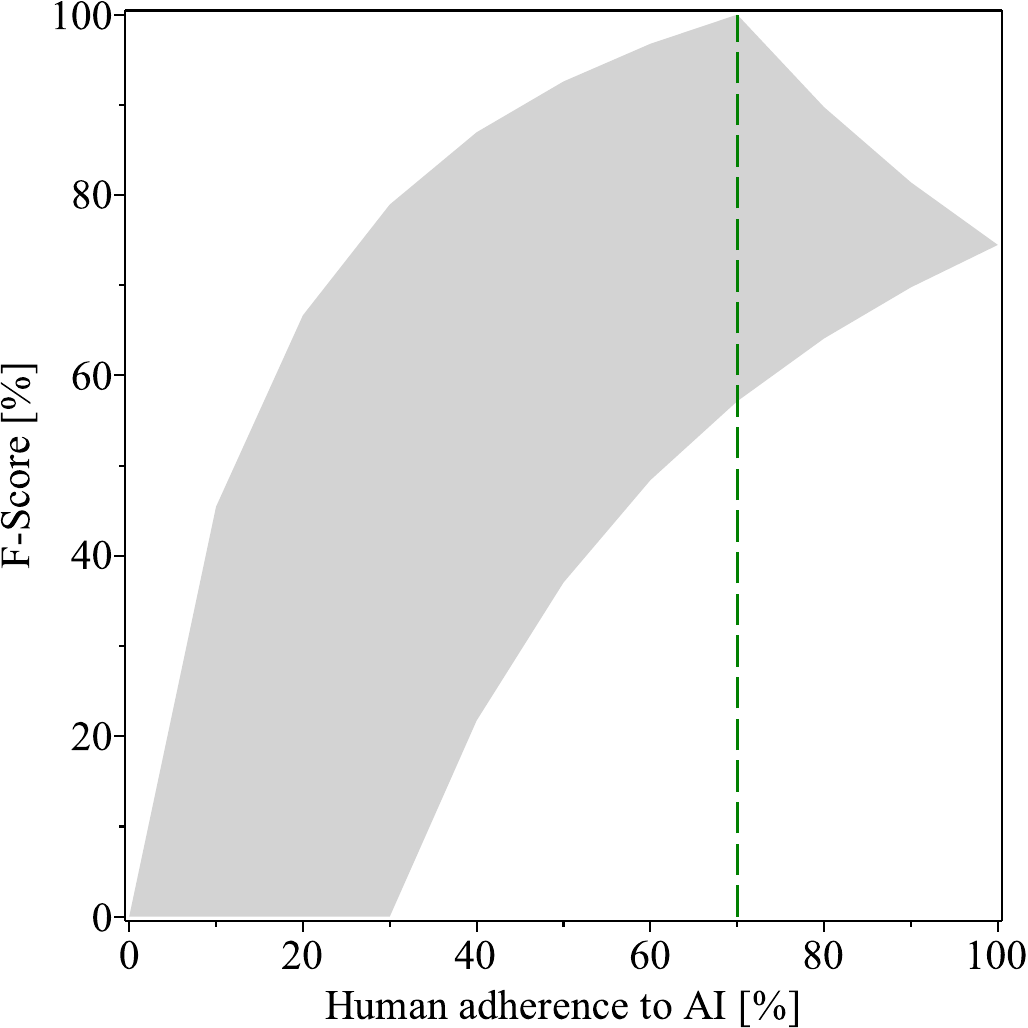}
        (b) $F_{0.5}$-Score
    \end{minipage}\hfill
\caption{The area of attainable (a) $F_{1}$-scores and (b) $F_{0.5}$-scores for a given AI accuracy of 70\% and different levels of human adherence.}
\label{fig:f-scores}
\end{figure}

With a level of adherence that is strictly lower than 100\%, we again observe intervals that represent all attainable levels of decision quality.
For instance, for an adherence level of $\mathcal{A}=70\%$, we obtain $F_1 \in [57.14\%, 100\%]$.
Thus, as with the assessment based on the accuracy metric, the best possible $F_1$-score can be obtained when the adherence level equals the accuracy of the AI system---making our conceptualization consistent across different metrics of decision quality.
However, in contrast to the assessment based on accuracy, we observe non-linear boundaries of the area encompassing all attainable levels of decision quality resulting from the assessment via the $F_1$-score. 
For instance, for an adherence level of $\mathcal{A}=60\%$, we have $F_1 \in [46.15\%, 92.31\%]$, and for an adherence level of $\mathcal{A}=50\%$, we obtain $F_1 \in [33.33\%, 83.33\%]$.
If the boundaries were linear, given the intervals for $\mathcal{A}=50\%$ and $\mathcal{A}=70\%$, we would expect $F_1 \in [45.24\%, 91.67\%]$ for $\mathcal{A}=60\%$, which is in contrast to the actual calculated values $[46.15\%, 92.31\%]$.
We visualize these non-linearities resulting from the assessment of decision quality based on the $F_1$-score in \Cref{fig:f-scores}~(a).

Additionally, in \Cref{fig:f-scores}~(b), we depict the attainable decision quality in terms of the $F_{0.5}$-score, which puts more weight on the precision than the recall (i.e., cases where false positives are considered worse than false negatives).
Thus, the $F_{0.5}$-score would consider adhering to wrong AI recommendations more problematic than overriding correct ones.
We observe that based on the unequal weighting of precision and recall, the shape of the boundaries changes compared to the $F_{1}$-score, with wider intervals of attainable decision quality.

\section{Discussion and Conclusion}\label{sec:conclusion}

\paragraph{Summary of Findings and Implications}
In this work, we study the relationship between reliance behavior and decision quality in AI-assisted decision-making.
We show that any given \textit{quantity} of human adherence to AI recommendations is associated with a specific range of attainable decision-making accuracy, depending on the \textit{quality} of reliance, i.e., the ability of humans to adhere to AI recommendations if and only if they are correct.
Vice versa, we also show that any accuracy level can be achieved through fundamentally different reliance behavior.
This has implications for assessing the effectiveness of interventions, such as explanations, in AI-assisted decision-making. 
In particular, our work highlights the importance of disentangling and assessing \textit{both} effects on accuracy \textit{and} reliance behavior in order to derive meaningful implications on how interventions affect decision-making.
For instance, by not capturing effects on reliance, we may conclude that an intervention led to an increase in accuracy, without understanding that this increase was driven solely by human over-reliance on AI advice.

We also characterize the conditions under which human-AI complementarity is achievable.
In particular, we show that under- and over-reliance\footnote{Recall that we define \textit{under-reliance} globally as $\mathcal{A}<Acc_{AI}$, and \textit{over-reliance} as $\mathcal{A}>Acc_{AI}$.} are not symmetrical regarding their implications for complementarity.
While complementarity is possible in the case of over-reliance, there is no hope when the human-in-the-loop under-relies past a certain threshold.
Notably, this threshold may be very high when the AI performs well.
For instance, at an AI accuracy of $90\%$, any adherence to AI recommendations of less than 80\% can \textit{never} lead to a decision-making accuracy that is better than the AI baseline.
Especially when the human-in-the-loop is not aware of such high AI performance, it might be unrealistic to expect complementarity.
Conversely, our work demonstrates that complementarity can still occur in empirical studies even when humans cannot distinguish correct from incorrect AI recommendations---and the likelihood of this can be significant.
This should be considered when interpreting empirical findings.

Finally, we propose a visual framework and demonstrate its practical relevance by applying it to prior empirical studies.
Our framework manages to capture a range of previous findings within a single visualization and extends them by additional observations.
For instance, it offers helpful additional information by disentangling reliance behavior in its basic components of correct/wrong adherence and overriding.
The proposed framework also allows us to compare empirical findings across studies by contrasting the human ability to distinguish correct from wrong AI recommendations.
Taken together, our work offers a blueprint and helpful metrics to guide the evaluation and design of effective interventions for AI-assisted decision-making.

\paragraph{Limitations and Future Work}
We acknowledge several assumptions that limit the scope of this work and allude to different areas that merit follow-up work.
First, some of the key theoretical results in our study hinge on the assumption of a binary decision-making task.
Although many critical decisions in the real world can be framed as binary, such as lending, hiring, or predicting recidivism, we recognize that certain tasks may involve more than two decision alternatives or even resemble regressions---e.g., in situations where a bank agent must decide on specific loan amounts to issue.
While we demonstrate that our visual representation of the interplay between reliance and decision quality remains useful for interpreting empirical results in multi-class decision-making, a logical extension would be to generalize our theoretical findings to cases with more than two decision-making alternatives.
In such scenarios, adjustments to our reliance taxonomy would be necessary to address situations where overriding a mistaken AI recommendation could still result in an incorrect decision.
Moreover, extending our framework to regression tasks may warrant further exploration, albeit in a more comprehensive manner.

Additionally, the evaluation of reliance behavior can be either sequential or concurrent, meaning that a human-in-the-loop may sometimes make an initial guess before receiving AI advice (sequential) or make the final decision concurrently with receiving AI advice (concurrent), as discussed by \shortciteA{tejeda2022ai}.
We focus on the concurrent paradigm in this work, recognizing that sequential setups require a different conceptualization of reliance, as introduced, e.g., by~\shortciteA{schemmer2023appropriate}.
As such, our work is complementing related literature on sequential AI-assisted decision-making, such as the works by \shortciteA{dunning2023humans} or \shortciteA{van2013framework}.
Finally, the decision-making setup that we consider is such that a human is relying on AI recommendations, i.e., either adhering to or overriding them.
To this point, it could be interesting to build on previous research on ``artificial trust''~\shortcite{azevedo2021unified} and reverse this relationship between human and AI agent, such that AI agents are relying on human input.

It is also important to note that many of our theoretical results rely on the assumption that AI recommendations are more accurate than chance.
This is motivated by prior work that has shown that unreliable automation may be worse than no automation at all~\shortcite{wickens2007benefits}; and an AI system that performs worse or at most as good as random guessing is an extreme case of unreliability.
While we consider it a reasonable assumption for AI systems in deployment to perform better than chance, we acknowledge that in certain cases, this cannot be guaranteed ex ante.

Our visual framework is also limited in its ability to compare empirical findings across studies with different AI accuracy.
Expanding it to account for varying AI accuracy would necessitate a three-dimensional visualization, introducing a third axis for $Acc_{AI}$, a task we defer to future work. 
Lastly, although it is often convenient to attribute initial recommendations to an AI system, such as a supervised machine learning model, our work makes no assumptions about the specific origin of these recommendations.
They could originate from any form of technology, recommender system, or even another human being.

\vskip 0.2in
\bibliography{refs}
\bibliographystyle{theapa}

\end{document}